\documentclass[journal,12pt,onecolumn,draftclsnofoot]{IEEEtran}

    \IEEEoverridecommandlockouts

    \ifCLASSINFOpdf
    
    \else
    
    \fi
    
    \usepackage{booktabs}
    \usepackage{enumitem}

    \usepackage{bbm}
    \usepackage{bigints}
    \usepackage{amsmath}
    \usepackage{amssymb}
    \usepackage{outlines}
    \usepackage{array}
    \usepackage{float}
    \usepackage{kotex}
    \usepackage{graphicx}
    \usepackage{bm}

    \usepackage[font=footnotesize]{caption}
    \usepackage{algorithm}
    \usepackage{algpseudocode}
    \usepackage{chngcntr}
    \usepackage{csquotes}
    \usepackage{cite}
    \usepackage{multicol}
    \usepackage{stmaryrd}
    \usepackage{tabularx,booktabs}
    \usepackage{xcolor}
    \newcolumntype{C}{>{\centering\arraybackslash}X} 
    \usepackage{lipsum} 
    \usepackage{multirow}
    \usepackage{caption,subcaption}

    \newsavebox{\measurebox}
    \DeclareMathOperator*{\argmax}{arg\,max}

    \newtheorem{definition}{Definition}

\begin{document}
    %
    \title{Test-Time Scalable AI-RAN: Inference Time Allocation for Cell-Free MIMO}
    %
    %
    %
    
    \author{Seonghoon~Yoo,~\IEEEmembership{Graduate Student Member,~IEEE,} Sangwoo~Park,~\IEEEmembership{Member,~IEEE,} \\Seok-Hwan~Park,~\IEEEmembership{Senior Member,~IEEE,} and
    Joonhyuk~Kang,~\IEEEmembership{Member,~IEEE}
    
    \thanks{Seonghoon Yoo and Joonhyuk Kang are with the School of Electrical Engineering, Korea Advanced Institute of Science and Technology (KAIST), Daejeon 34141, South Korea (e-mail: shyoo902@kaist.ac.kr, jhkang@ee.kaist.ac.kr).}
    \thanks{Sangwoo Park is with  the Department of Engineering, King’s College London, WC2R
2LS London, U.K. (e-mail: sangwoo.park@kcl.ac.uk).}
    \thanks{Seok-Hwan Park is with the School of Electrical Engineering, Hanyang University ERICA, Ansan 15588, South Korea (e-mail: seokhwanpark@hanyang.ac.kr).}
    
    }
    
    \markboth{IEEE XXX,~Vol.~xx, No.~xx, XXX~2026}%
    {Yoo \MakeLowercase{\textit{et al.}}: Test-Time Scalable AI-RAN: Inference Time Allocation for Cell-Free MIMO}
    
    \maketitle
    
\begin{abstract}
Artificial intelligence-enabled radio access networks (AI-RANs) are envisioned to consist of multiple AI-based modules, potentially developed independently by different vendors. In this work, we study AI-RAN-enabled cell-free MIMO systems, with a particular focus on the system implications of modern AI models.  Specifically, we focus on the phenomenon of \emph{test-time scalability} popularized by large language models (LLMs), under which model performance improves as additional computational resources are allocated at testing time. By noting that the optimal amount of additional computational resources for each AI module should in general depend on its interaction with the other modules as well as with the underlying wireless channels, we propose a generic framework that enables optimal resource allocation for each test-time scalable module in cell-free MIMO systems. Experimental results demonstrate the effectiveness of the proposed framework in fully exploiting the potential of test-time scalable AI-RANs in cell-free MIMO systems.
\end{abstract}
    

    \IEEEpeerreviewmaketitle

    \section{Introduction}
    \label{sec:intro}
    
\subsection{Context and Motivation}

\IEEEPARstart{T}{he} number of user equipments (UEs) and their capacity demands are expected to grow rapidly in next-generation wireless networks \cite{Ngo2024largeUE, 3gpp_AI_RAN}. Cell-free multiple-input multiple-output (MIMO) is a promising architecture to meet these demands, where many geographically separated radio units (RUs) jointly serve UEs over a wide area
\cite{Bjorson2020scale_CFMIMO}. By coordinating these RUs through a central processor (CP), cell-free massive MIMO can improve coverage and spectral efficiency compared with conventional cell-based systems. Such joint management, however, inevitably requires communication between the CP and RUs over capacity-limited fronthaul links.

In this setting, functional split between the CP and RUs becomes a key design factor, as it determines the tradeoff among downlink performance, computational burden, processing latency, and fronthaul overhead. A representative split is \textit{Option~7.2}, where high physical-layer (PHY)
baseband processing such as channel coding, modulation, and precoding is
handled at the CP side, while low-PHY processing such as FFT/IFFT operations
and radio-frequency (RF) functions such as up/down-conversion and power
amplification remain at the RU side 
\cite{Larsen2019functional_split,Bjornson2024ORAN_cellfree}. One notable example of \emph{Option 7.2} is \emph{compress-and-precode} \cite{liu2016joint, liu2017fronthaul, khorsandmanesh2023fronthaul}, where the CP shares the compressed precoding matrix as well as the UE data streams with the RUs, and each RU locally
performs precoding and the radio transmission accordingly. However, such \emph{data sharing} nature of compress-and-precode functional split easily makes the fronthaul overhead overwhelming as the number of users rapidly grows \cite{Seokhwan2023precoder_fronthaul, sangwoo2025scalable}.   

In this work, we focus on \emph{precode-and-compress} transmission, where the
CP directly sends
the compressed precoded signal to the RUs 
\cite{park2013joint, liu2017fronthaul}. This
architecture simplifies RU-side processing and avoids UE-data sharing over the
fronthaul, since each RU only receives the baseband signal to be transmitted
through its antennas rather than the data streams of all UEs
\cite{Bjorson2020scale_CFMIMO,Erik2020Precoding_CFMIMO}. However, under
limited fronthaul capacity, precode-and-compress can suffer from a severe rate
loss unless fronthaul quantization is carefully designed together with
precoding.

To address the high-complexity challenges, artificial intelligence (AI)-based approaches have been studied for cell-free MIMO and precode-and-compress transmission. For example, meta-learning-based dimension reduction of the precoded signal was proposed in~\cite{YuWei2024Fronthaul}, while neural-network-based multivariate quantization was studied in~\cite{sangwoo2025scalable} (see Sec.~\ref{subsec:related_work} for further related works). These studies suggest that AI-based modules can help bring the cooperative gain of cell-free massive MIMO closer to practical deployment under fronthaul and computation constraints. This direction is also aligned with AI-radio access network (AI-RAN) and open
RAN (O-RAN) paradigms for open and intelligent sixth-generation
(6G) RAN, where RAN functions can be implemented as modular AI applications over open and programmable interfaces \cite{3gpp_AI_RAN,Yoo2026AI_RAN,Polese2024O-RAN}. Such modular operation is particularly relevant in open and AI-native RAN
architectures, where independently developed AI functions can coexist through
standardized interfaces and jointly affect the network performance \cite{Hoffmann2024ORAN}.

In recent years, large language models (LLMs) have driven remarkable breakthroughs in AI, motivating their adoption in wireless networks \cite{qiao2025todma, liang2026large}. In this work, we focus on the phenomenon of \emph{test-time scalability}, which has been popularized by LLMs. Test-time scalability \cite{snell2024scaling, ellis2025theory, zhang2025test_time_generalization} is a property often observed in modern AIs under which the performance of the underlying model improves as additional computational resources are allocated at testing time  \cite{ramesh2026test,huang2017snapshot,  tao2026reliable, hegde2024calibrated}.


Specifically, we focus on the optimal
utilization of multiple separately designed test-time scalable AI modules to enhance the end-to-end performance of the underlying cell-free MIMO system. We first describe the ways to achieve joint test-time scalability of separately designed AI modules, followed by optimally choosing the corresponding inference time of each AI module. Such inference time allocation allows one to bring the maximal gain of AI-based modules for any cell-free MIMO setting of interest. 

    
\subsection{Related Work}\label{subsec:related_work}

Related works can be categorized into three main directions:
(\emph{i}) precoding and fronthaul design for cell-free massive MIMO;
(\emph{ii}) AI-assisted fronthaul quantization and precode-and-compress transmission;
and (\emph{iii}) AI-native RAN architectures and test-time scalable AI.

Cell-free massive MIMO has been studied in terms of scalable operation, downlink precoding, fronthaul compression, and their joint design. Scalable operation has been investigated by limiting cooperation, signal processing, and fronthaul signaling as the network size grows~\cite{Bjorson2020scale_CFMIMO}. Precoding methods have been developed to improve spectral efficiency and
interference management in cell-free transmission \cite{Erik2020Precoding_CFMIMO,Feng2022WMMSE}. Fronthaul compression,
quantization, and bit-allocation strategies have also been studied for capacity-limited fronthaul links
\cite{YuWei2024Fronthaul}, and joint precoding and fronthaul design has been considered for end-to-end downlink optimization~\cite{Seokhwan2023precoder_fronthaul}. These works, however, typically assume fixed online operation and do not explicitly address how a limited test-time budget should be allocated across different design modules.

AI-assisted techniques have also been applied to cell-free massive MIMO and precode-and-compress transmission. Deep learning has been used for tasks such as channel estimation, while learning-based fronthaul quantization has been investigated to reduce the complexity of compressed precoded signal design \cite{YuWei2024Fronthaul,sangwoo2025scalable}. In parallel, AI-RAN and O-RAN have provided an architectural basis for deploying AI or machine learning (ML) modules over open, cloud-native, and multi-vendor RAN infrastructures \cite{3gpp_AI_RAN,Polese2024O-RAN}.  For instance, real-time AI-enabled RAN functions have been demonstrated at the PHY layer \cite{Melodia2025AI-RAN_PHY} while the tradeoff between AI-based service quality and latency in decentralized setting under practical user mobility has been studied in \cite{zhang2026decentralized}. These developments motivate the question of how separately designed AI modules should share limited test-time resources when their outputs jointly determine the downlink performance.
    
\subsection{Main Contribution}

The main contributions of this work can be summarized as follows:
\begin{itemize}
\item We formulate an optimal inference time allocation problem for
test-time scalable AI-RAN in cell-free MIMO under the precode-and-compress
downlink architecture with capacity-limited fronthaul links.

\item We construct AI-based precoding and fronthaul quantization modules.
The precoder module is based on AI-assisted weighted minimum mean-square
error (WMMSE) precoding, while the quantizer module is based on deep
task-based quantization \cite{Shlezinger2021Deep_taskbased_quantizer}
with scalar uniform quantizers.

\item We develop test-time scalable precoder and quantizer modules based
on parallel sampling \cite{brown2024monkey_parallelsampling}. Under a given inference time budget, the modules
generate stochastic candidate precoders and quantization-dequantization
function tuples, and the final pair is selected according to the
finite-blocklength sum-rate metric.

\item We optimize the inference time allocation between the scalable
precoder and quantizer by accounting for the coherence block time-frame
structure. Simulation results validate the test-time scalability of the
proposed modules and show that the proposed allocation improves the
finite-blocklength sum-rate over the considered baselines.
\end{itemize}

The remainder of this paper is organized as follows. Sec.~II describes the considered cell-free MIMO downlink system and problem formulation. Sec.~III introduces the non-scalable AI-based precoding and fronthaul quantization modules, and Sec.~IV develops their test-time scalable counterparts. Sec.~V presents the joint candidate selection and inference time allocation. Simulation results are provided in Sec.~VI, followed by the conclusion in Sec.~VII.

    \section{System Model}
    \begin{figure}[t]
        \centering
        \includegraphics[width=.75\linewidth]{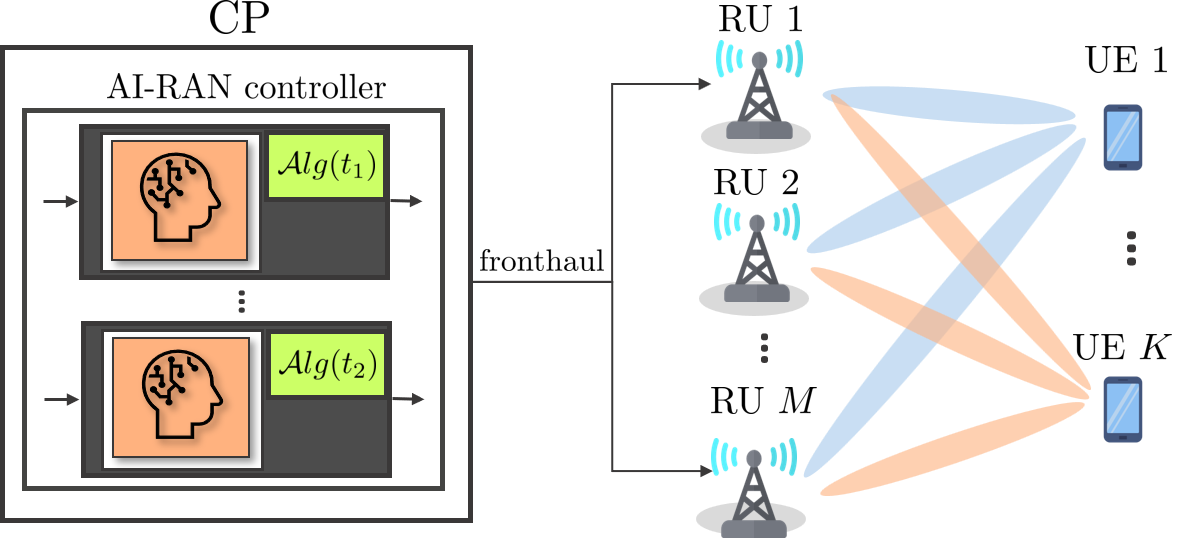}
        \caption{System model of the considered AI-RAN-enabled cell-free MIMO architecture, consisting of a single CP, $M$ RUs, and $K$ UEs. The CP hosts an AI-RAN controller that runs algorithms based on multiple AI apps for baseband processing, and transmits compressed baseband signals to the RUs over capacity-limited fronthaul links.}
        \label{system_model}
    \end{figure}
    As illustrated in Fig. \ref{system_model}, we consider a cell-free MIMO downlink system with $K$ UEs and $M$ distributed RUs, whose coordination is carried out by a single CP. We focus throughout the paper on the precode-and-compress functional split \cite{park2013joint, liu2017fronthaul}. 
    
    
    
    \subsection{Setting}
    \label{Setting}
    Let us denote as $N$ the number of transmit antennas for each RU and further assume that each UE has a single receive antenna. We will assume that the fronthaul link has a limited capacity of $C_F$ bit/sec, and that the bandwidth of the radio interface is $B$ Hz. Accordingly, we set the sample period as $T_s=1/B$~sec, allowing the system to transmit $L=\lfloor T/T_s \rfloor$ number of symbol vectors $(\mathbf{s}_i)_{i=1}^L$ with $\mathbf{s}_i=[s_{i,1},...,s_{i,K}]^\top \overset{\text{i.i.d.}}{\sim} P_\mathbf{s}$ for $i=1,...,L$ given some fixed distribution $P_\mathbf{s}$  during the coherence time of $T$~sec. We further assume $\mathbb{E}[\mathbf{s}_i\mathbf{s}_i^H]=\mathbf{I}_K$ in order to ensure that the
data streams are normalized to have unit power. 

During this coherence time, the channel between the RUs and the UEs is assumed
to be constant and hence can be represented as a single matrix
$\mathbf{H} \in \mathcal{H}=\mathbb{C}^{K \times NM}$ whenever the effective
channel admits frequency-flat fading. We assume this flat-fading property
throughout this work, which can be justified by either choosing a sufficiently
narrow bandwidth $B$ or considering each orthogonal frequency-division
multiplexing (OFDM) subcarrier separately. The aggregate downlink channel state information (CSI) $\mathbf{H}$ is assumed to be available at the CP at the beginning of each coherence interval for precoder and quantizer design \cite{park2013joint}.
    

    In order to convey as much information as possible in the symbols
$(\mathbf{s}_i)_{i=1}^{L}$, the CP designs a precoding matrix
$\mathbf{W}\in\mathbb{C}^{NM\times K}$ based on the channel matrix
$\mathbf{H}$. Let $\mathbf{h}_k\in\mathbb{C}^{NM}$ denote the aggregate channel
vector from all RUs to UE $k$, so that the $k$-th row of $\mathbf{H}$ is
$\mathbf{h}_k^H$. We write
$\mathbf{W}=[\mathbf{w}_1,\ldots,\mathbf{w}_K]$, where
$\mathbf{w}_k\in\mathbb{C}^{NM}$ denotes the aggregate precoding vector for UE
$k$, and let $[\mathbf{W}]_m\in\mathbb{C}^{N\times K}$ denote the submatrix of
$\mathbf{W}$ corresponding to the $m$-th RU. Under the precode-and-compress
functional split, the CP then computes the precoded symbol vector as
$\mathbf{x}_i=\mathbf{W}\mathbf{s}_i$ for $i=1,\ldots,L$. Specifically, the $i$-th quantized precoded symbol vector $\tilde{\mathbf{q}}_i \in \{0,1\}^b$ is obtained by CP via
    \begin{equation}
        \tilde{\mathbf{q}}_i = Q(\mathbf{x}_i),
        \label{eq:quantized_signal}
    \end{equation}
    using the pre-determined quantization function   $Q:\mathbb{C}^{NM}\rightarrow\{0,1\}^{b}$ where the bitwidth $b=\lfloor C_F/B\rfloor$ is given by the fronthaul capacity $C_F$ and the system bandwidth $B$. 
    
    The $b$ bits are then partitioned into $M$ parts as $\tilde{\mathbf{q}}_i = [\tilde{\mathbf{q}}_{i}^1,...,\tilde{\mathbf{q}}_{i}^M]$, with only the $m$-th part $\tilde{\mathbf{q}}_i^m \in \{0,1\}^{b_m}$ being  transmitted to the $m$-th RU. Note that $\sum_{m=1}^M b_m=b$. Upon the reception of $b_m$ bits through the fronthaul link, the $m$-th RU reconstructs its dedicated symbol vector by applying its own pre-determined dequantization function $Q_m^{-1}:\{0,1\}^{b_m}\rightarrow \mathbb{C}^{N}$ as 
    \begin{equation}
    \label{eq:dequantized_signal}
        \tilde{\mathbf{x}}_i^{m} = Q_{m}^{-1}(\tilde{\mathbf{q}}_{i}^{m}).
    \end{equation}
    The dequantization function is designed such that the reconstructed signal
satisfies the per-RU average transmit-power constraint
$\mathbb{E}[\|\tilde{\mathbf{x}}_i^m\|^2]\leq P_{\rm tx}$, where $P_{\rm tx}$ is the maximum
transmit power of each RU.
    The $m$-th RU then finally transmits $\tilde{\mathbf{x}}_i^{m} $ through its own $N$ antennas: Denoting as $\mathbf{H}_m \in \mathbb{C}^{K \times N}$ the channel matrix between $m$-th RU and the $K$ UEs, the received signal $\mathbf{y}_i \in \mathbb{C}^{K\times 1}$ at the UE side can be accordingly represented as \begin{equation}\label{eq:overall_mapping}
        \mathbf{y}_i = \sum_{m=1}^{M}  \mathbf{H}_{m} \tilde{\mathbf{x}}_i^{m} + \mathbf{n}_i,
    \end{equation}
    where $\mathbf{n}_i \sim \mathcal{CN}(\mathbf{0},\sigma^2\mathbf{I}_{K})$ is the additive white Gaussian noise with some fixed variance $\sigma^2$.

    Although information theory suggests that the precoding matrix $\mathbf{W}$, quantization function $Q$, and the dequantization functions $\{Q_m^{-1}\}_{m=1}^M$ should be jointly optimized in order to convey the maximal information within the symbols $(\mathbf{s}_i)_{i=1}^L$, due to the coupling between the quantization and dequantization functions that inevitably introduces a signaling overhead between the CP and the RUs, it is often a common practice to update the quantization functions only occasionally. In the next subsection, we formalize this joint optimization perspective more generally.

    \subsection{Goal}
    \label{subsec:goal}
    As mentioned above, joint optimization of the precoding matrix $\mathbf{W}$ and the quantization-dequantization functions $ Q(\cdot),\, \{ Q_m^{-1}(\cdot) \}_{m=1}^{M}$ should take into account both (\emph{i}) computation overhead as well as (\emph{ii}) signaling overhead. To this end, we formalize the joint optimization problem \emph{under the fixed computation/signaling overhead} as 
\begin{equation}
\label{eq:objective}
\begin{aligned}
&\begin{aligned}
\max_{\mathbf{W},\,Q(\cdot),\,\{Q_m^{-1}(\cdot)\}_{m=1}^{M}}
\quad &
T_{\mathrm{prep}}
R_{\epsilon}\!\left(
T_{\mathrm{prep}};
P_{\mathbf{y}^{\mathrm{outdated}}\mid\mathbf{s}}
\right)
\\
&\quad+
(T-T_{\mathrm{prep}})
R_{\epsilon}\!\left(
T-T_{\mathrm{prep}};
P_{\mathbf{y}\mid\mathbf{s}}
\right)
\end{aligned}
\\[2pt]
&\hspace{2em}
\begin{aligned}
\mathrm{s.t.}\quad &
T_{\mathrm{prep}}=t_1+t_2+t_3,
\\
&
\mathbf{W}
\in
\mathcal{A}\mathrm{lg}_{\mathrm{precode}}(t_1),
\\
&
\left\{
Q(\cdot),
\{Q_m^{-1}(\cdot)\}_{m=1}^{M}
\right\}
\in
\mathcal{A}\mathrm{lg}_{\mathrm{comp}}(t_2),
\\
&
\sum_{m=1}^{M}
\log
\left|
Q_m^{-1}(\cdot)
-
\bar{Q}_m^{-1}(\cdot)
\right|
\leq C_F t_3,
\end{aligned}
\end{aligned}
\end{equation}
where $R_\epsilon(t;P_{\mathbf{y}|\mathbf{s}})$ is used to denote the maximal number of information bits that can be \emph{reliably} conveyed during one channel use -- from $\mathbf{s}$ to $\mathbf{y}$ under the conditional distribution $P_{\mathbf{y}|\mathbf{s}}$ -- whose  reliability is jointly determined in a block-wise manner with the block size being $\lfloor t/T_s\rfloor$.  Here, information is said to be reliably conveyed if the probability of error is no larger than $\epsilon$ where an error occurs whenever the block-wise decoded message with $\lfloor t/T_s\rfloor$ channel uses differs from the intended message at the transmitter side.


In (\ref{eq:objective}), the conditional distribution $P_{\mathbf{y}|\mathbf{s}}$ is given by (\ref{eq:quantized_signal})--(\ref{eq:overall_mapping}) that depends on the optimization variables $\mathbf{W}, Q(\cdot), \{ Q_m^{-1}(\cdot)\}_{m=1}^M $. The \emph{outdated} conditional distribution $P_{\mathbf{y}^\text{outdated}|\mathbf{s}}$ is also given by (\ref{eq:quantized_signal})--(\ref{eq:overall_mapping}), but with a stark difference that $P_{\mathbf{y}^\text{outdated}|\mathbf{s}}$ is independent of the optimization variables but to use outdated precoding matrix and outdated quantization functions (e.g., the ones obtained in the previous coherence block). 

\begin{figure}[t]
    \centering
    \includegraphics[width=.75\linewidth]{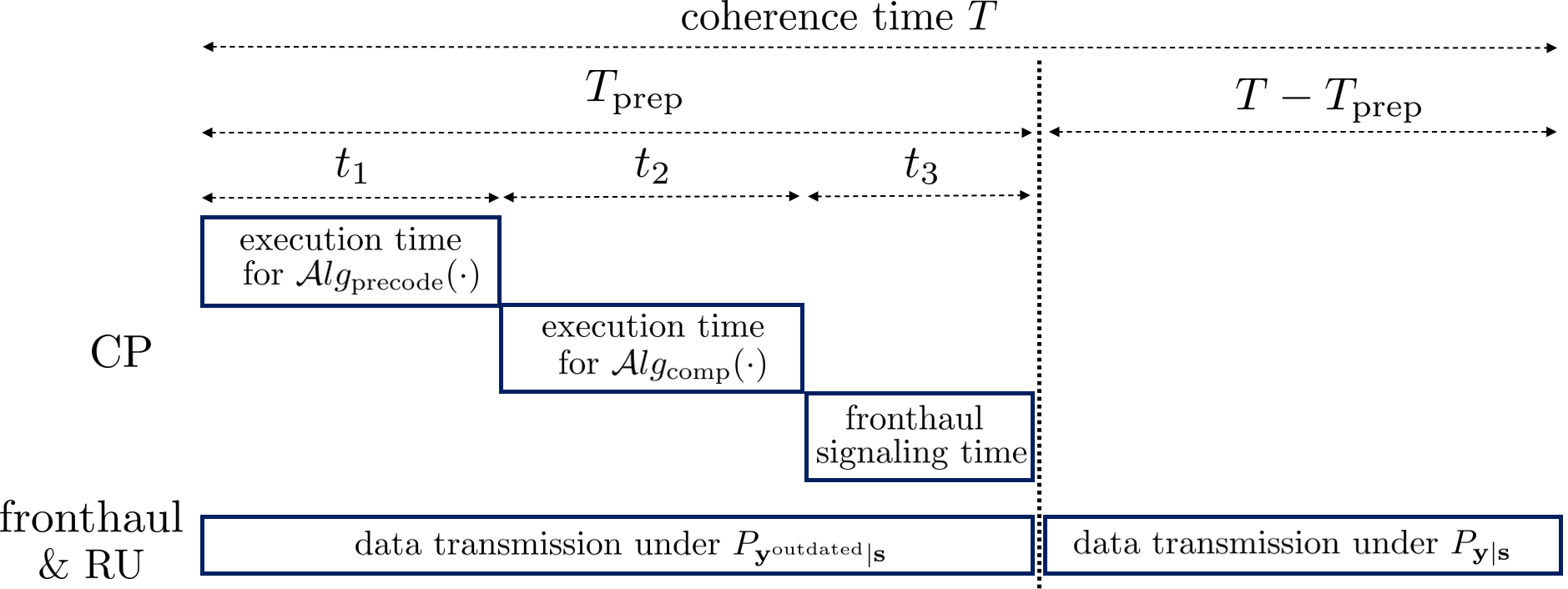}
   \caption{Time frame of the proposed AI-RAN transmission within a coherence block
of duration $T$. During the preparation interval
$T_{\rm prep}=t_1+t_2+t_3$, transmission uses the precoder and
quantization-dequantization functions from the previous coherence block; after
the preparation interval, transmission uses the current coherence block design.}
\label{time_frame}
\end{figure}

    Furthermore, in (\ref{eq:objective}), $\mathcal{A}\textrm{lg}_\text{precode}(t)$ is used to denote the class of all algorithms that yield feasible precoding matrix and can be run within the execution time of $t$~sec; similarly,  $\mathcal{A}\textrm{lg}_\text{comp}(t)$ is the class of all algorithms that yield feasible quantization-dequantization functions that can be run within the execution time of $t$~sec. As illustrated in Fig.~\ref{time_frame},   $t_1$ denotes the execution time used for designing the precoding matrix $\mathbf{W}$, $t_2$ denotes the execution time used for designing the quantization and dequantization functions $\{Q(\cdot), \{Q_m^{-1}\}_{m=1}^M\}$, and $t_3$ denotes the fronthaul signaling time used for updating the dequantization functions $\{Q_m^{-1}(\cdot)\}_{m=1}^M$ at the RUs. We have also considered a \emph{base} dequantization  $\bar{Q}_m^{-1}(\cdot)$ for each of the $m$-th RU, so that only the difference with respect to the base function needs to be signaled through the fronthaul link. Such difference has been expressed using an operator $\log |f(\cdot)|$ that counts the number of bits required to fully express some arbitrary function $f$.

    Note that in (\ref{eq:objective}), the time required for matrix-vector multiplication to get  $\mathbf{x}_i = \mathbf{W}\mathbf{s}_i$ as well as the time required to actually run the quantization/dequantization functions are ignored as these operations can be carried out within the time window of the preceding sample period \cite{Melodia2025AI-RAN_PHY, Vieira2014real_time_testbed}.

    In most of the literature \cite{Erik2020Precoding_CFMIMO, Seokhwan2023precoder_fronthaul, Vieira2014real_time_testbed}, $t_1, t_2, t_3$ are fixed in advance and the goal is to design the algorithms $\mathcal{A}\textrm{lg}_\text{precode}$ and $\mathcal{A}\textrm{lg}_\text{comp}$ in order to convey as much information as possible within the symbols $(\mathbf{s}_i)_{i=1}^L$. In contrast, we take a different viewpoint: we adopt (existing, well-established) test-time scalable AI modules for $\mathcal{A}\textrm{lg}_\text{precode}$ and $\mathcal{A}\textrm{lg}_\text{comp}$ and focus on optimizing $t_1,t_2,t_3$. 
    


    We first describe in the sequel the conventional AI modules for precoding and quantization, followed by their natural extension that enables test-time scalability.

\section{Conventional Non-Scalable AI-Based Modules}
\label{sec:non_scalable_ai_modules}

In this section, we describe our choice of conventional (non-scalable) AI-based precoding module as well as AI-based quantization/dequantization modules.


\subsection{AI-Based Precoder Design}
\label{subsec:non_scalable_ai_precoder}

We first describe AI-based precoding. Although there exist many other AI-driven approaches \cite{Huang2019DL_precoding, Zhang2022DL_precoding}, we focus in this work on the particular type of \emph{model-based learning} \cite{shlezinger2023model}, often referred to also as \emph{deep unfolding/unrolling} \cite{hershey2014deepunfolding}. We will consider the design of AI-based precoding by replacing a few sub-modules of WMMSE algorithm \cite{Shi2011WMMSE, Feng2022WMMSE} with AI-driven modules, similarly to \cite{pellaco2020wmmse_deepunfolding, Hu2021deep_unfloding}. In particular, we focus on AI-based data-driven initialization of WMMSE algorithm given its well-known importance on determining the optimality of its convergent solution. We start by briefly reviewing the WMMSE algorithm.

\subsubsection{WMMSE algorithm}
The WMMSE algorithm \cite{Shi2011WMMSE} yields a precoding matrix
$\mathbf{W}\in\mathcal{W}$ satisfying the per-RU transmit-power constraint,
i.e.,
\begin{equation}
    \mathcal{W}
    =
    \left\{
    \mathbf{W}\in\mathbb{C}^{NM\times K}:
    \|[\mathbf{W}]_m\|_F^2 \leq P_{\rm tx},\;
    m=1,\ldots,M
    \right\}.
    \label{eq:wmmse_feasible_set}
\end{equation}

The WMMSE algorithm alternately updates the three variables:  \emph{MMSE receive filters}
$\{u_k\}_{k=1}^{K}$, \emph{weights} $\{\omega_k\}_{k=1}^{K}$, and the precoding
matrix $\mathbf{W}$. Specifically, given the precoder $\mathbf{W}$, the MMSE receive filter
of UE $k$ is determined as
\begin{equation}
    u_k
    =
    \frac{\mathbf{h}_k^H\mathbf{w}_k}
    {\sum_{\ell=1}^{K}
    |\mathbf{h}_k^H\mathbf{w}_{\ell}|^2+\sigma^2},
    \label{eq:wmmse_receiver_update}
\end{equation}
for $k=1,...,K$.
The weight for the UE $k$ is then updated as
\begin{equation}
    \omega_k =\frac{1}{1
    -2\Re\{u_k^*\mathbf{h}_k^H\mathbf{w}_k\}
    + |u_k|^2
    \left(
    \sum_{\ell=1}^{K}
    |\mathbf{h}_k^H\mathbf{w}_{\ell}|^2+\sigma^2
    \right)}  
    \label{eq:wmmse_weight_update}
\end{equation}
for $k=1,...,K$.
Given $\{{u_k}\}_{k=1}^{K}$ and $\{\omega_k\}_{k=1}^{K}$, the precoder is then
updated as
\begin{equation}
\begin{aligned}
[\mathbf{W}]_m
=&
\left(
\sum_{k=1}^{K}
\omega_k |u_k|^2 \mathbf{h}_{m,k}\mathbf{h}_{m,k}^H
+
\mu_m\mathbf{I}_N
\right)^{-1}
\\
&\times
\left[
\omega_1u_1\mathbf{h}_{m,1},\ldots,
\omega_Ku_K\mathbf{h}_{m,K}
\right]
\label{eq:wmmse_precoder_update}
\end{aligned}
\end{equation}
for $m=1,...,M$, where $\mu_m>0$ is chosen in order to ensure the per-RU transmit power constraint, which can be done via bisection search (see, e.g., \cite[Sec. IV-B]{Feng2022WMMSE} and \cite{Shi2011WMMSE}). The updates in \eqref{eq:wmmse_receiver_update}--\eqref{eq:wmmse_precoder_update} are repeated until convergence.

Note that the above alternating optimization requires an initialization for the precoding matrix, which has a substantial impact on the convergent solution \cite{Shi2011WMMSE}. In what follows, we define
$\mathrm{WMMSE}_{I_W}(\mathbf{W}_{\rm init},\mathbf{H})$ as the WMMSE algorithm
for the channel matrix $\mathbf{H}$ that repeats
\eqref{eq:wmmse_receiver_update}--\eqref{eq:wmmse_precoder_update} for $I_W$
iterations, with the precoder initialized as $\mathbf{W}_{\rm init}$. The full
description of $\mathrm{WMMSE}_{I_W}(\mathbf{W}_{\rm init},\mathbf{H})$ can be
found in Algorithm~\ref{alg:wmmse_refinement}.

\begin{algorithm}[t]
\caption{WMMSE-based precoding}
\label{alg:wmmse_refinement}
\begin{algorithmic}[1]
\State \textbf{Input: } Channel matrix $\mathbf{H}$, initial precoder $\mathbf{W}_{\rm init}$, number of WMMSE iterations $I_W$
\State Initialize $\mathbf{W}=\mathbf{W}_{\rm init}$
\For{$i=1,\ldots,I_W$}
    \State Update the receive filters $\{u_k\}_{k=1}^{K}$ using
    \eqref{eq:wmmse_receiver_update}
    \State Update the weights $\{\omega_k\}_{k=1}^{K}$ using
   \eqref{eq:wmmse_weight_update}
    \State Find the variables $\{\mu_m\}_{m=1}^{M}$ by a dual search
    \State Update the precoder $[\mathbf{W}]_m$ using
\eqref{eq:wmmse_precoder_update}
\EndFor
\State \Return $\mathbf{W}$
\end{algorithmic}
\end{algorithm}

\subsubsection{AI-assisted WMMSE algorithm}
Given the importance of the initialization of the WMMSE algorithm to improve
the quality of its convergent solution, we consider an AI-based, data-driven
initialization of the WMMSE algorithm in a manner similar to \cite{Hu2021deep_unfloding, pellaco2020wmmse_deepunfolding}.
Specifically, we first consider the following \emph{learnable,
channel-dependent initialization}
\begin{equation}
    g_W(\cdot;\theta_W):
    \mathcal{H}
    \rightarrow \mathcal{W},
    \label{eq:precoder_initialization_mapping}
\end{equation}
that takes as input the channel matrix $\mathbf{H}\in\mathcal{H}$ and outputs
the initial precoding matrix $\mathbf{W}_{\rm init}\in\mathcal{W}$, i.e.,
$\mathbf{W}_{\rm init}=g_W(\mathbf{H};\theta_W)$, where
$\theta_W\in\Theta$ is the learnable parameter vector within some  parameter
space $\Theta$.

In order to find an adequate parameter vector $\theta_W$ that aids the WMMSE
algorithm in obtaining a high-quality convergent solution, we optimize
$\theta_W$ across different channel realizations, i.e.,
$\mathbf{H}\sim P_{\mathbf{H}}$, where $P_{\mathbf{H}}$ denotes the channel
distribution. We first describe the ideal optimization problem assuming that
$P_{\mathbf{H}}$ is known, and then discuss the practical approximation under
the unknown $P_{\mathbf{H}}$. Specifically, we consider the following
optimization problem:
\begin{equation}
\label{eq:ai_wmmse_training_problem}
\begin{aligned}
\max_{\theta_W\in\Theta}\quad
&\mathbb{E}_{\mathbf{H}}
\left[
R_{{\rm p},\epsilon}(\mathbf{H},\mathbf{W})
\right]
\\
\text{s.t.}\quad
&\mathbf{W}
=
\mathrm{WMMSE}_{I_W}
\big(
g_W(\mathbf{H};\theta_W),\mathbf{H}
\big).
\end{aligned}
\end{equation}
where $R_{{\rm p},\epsilon}(\mathbf{H},\mathbf{W}) \equiv R_\epsilon(T; P_{\mathbf{y}^{C_F=\infty}|\mathbf{s}})$ whose conditional distribution $P_{\mathbf{y}^{C_F=\infty}|\mathbf{s}}$ is given by (\ref{eq:quantized_signal})--(\ref{eq:overall_mapping}) that adopts $\mathbf{W}$ as the precoding matrix under the infinite fronthaul capacity scenario (hence no quantization error). As briefly mentioned in Sec.~\ref{sec:intro}, we consider such an idealized fronthaul case to consider the independent design of AI modules from different vendors. We refer to Appendix~\ref{precoder_eval_appendix} for further details on the practical approximation of $R_{{\rm p},\epsilon}(\mathbf{H},\mathbf{W}) $ where the $M$ RUs adopt i.i.d. Gaussian codebooks and the $K$ UEs apply (mismatched) nearest-neighbor decoding rule \cite{scarlett2016dispersion, Choi2021FiniteSumrate}.


In practice, the distribution $P_{\mathbf{H}}$ is unknown hence one cannot
directly solve \eqref{eq:ai_wmmse_training_problem}. The common approach is to
collect a finite set of channel realizations and approximate the expectation in
\eqref{eq:ai_wmmse_training_problem} through the empirical averaging.


\subsubsection{Time complexity}
The WMMSE algorithm in Algorithm~\ref{alg:wmmse_refinement} requires
$I_W\cdot t_{\rm WMMSE}$~sec, where $t_{\rm WMMSE}$ is the hardware-specific
wall-clock time to run
\eqref{eq:wmmse_receiver_update}--\eqref{eq:wmmse_precoder_update}, whose
computational complexity is given by
$\mathcal{O}((NM)^3+K(NM)^2+K^2NM)$ \cite{Zhao2023complexity_wmmse}. The
AI-assisted WMMSE algorithm requires additional $t_{\rm init}$~sec, where
$t_{\rm init}$ is the hardware-specific wall-clock time to run $g_W$. For
instance, when adopting a multi-layer perceptron (MLP)  with $L_W$ hidden layers each with $d_W$ hidden
neurons to implement $g_W(\cdot;\theta_W)$, $t_{\rm init}$ will be dictated by
the corresponding inference computational complexity, which is of order
$\mathcal{O}(KNMd_W+L_Wd_W^2)$ \cite{simeone2022machine}. Therefore,
the precoder design time in \eqref{eq:objective} can be  modeled as
$t_1=t_{\rm init}+I_W\cdot t_{\rm WMMSE}$ for the AI-assisted case.

\subsection{AI-Based Quantizer Design}
\label{subsec:non_scalable_ai_quantizer}

We now consider AI-based quantizer design. Although vector quantization \cite{lee2016multivariate} and AI-based vector quantization \cite{sangwoo2025scalable} can also be used for fronthaul compression, we focus in this work on scalar uniform quantization by unveiling the potential of learnable task-oriented signal mapping \cite{Shlezinger2021Deep_taskbased_quantizer}. 


\subsubsection{Uniform quantizer module}
We first describe the standard, non-AI-based, approach for fronthaul compression using scalar uniform quantization. 
By noting  $\mathbf{x}_i \in \mathbb{C}^{NM}$, scalar uniform quantization needs to be separately applied to the $2NM$ scalars (real and imaginary part for each element), potentially with different scaling factors. In what follows, we will denote as $Q_\text{uni}: \mathbb{C}^{NM} \rightarrow \{0,1\}^b$ the quantization function implemented by the standard scalar uniform quantization; as $Q_{m, \text{uni}}^{-1}: \{0,1\}^{b_m}\rightarrow \mathbb{C}^N$ the corresponding dequantization function at the $m$-th RU. We refer to Appendix~\ref{quantizer_design_eval_appendix} for a more explicit description on the scalar uniform quantization functions.

\subsubsection{AI-assisted deep task-based quantization}
We next introduce AI-assisted deep task-based quantization. Unlike the standard
uniform quantizer, CP now first applies a learnable task-oriented signal mapping
before the uniform quantization. 

Specifically, given the precoded signal
$\mathbf{x}_i=\mathbf{W}\mathbf{s}_i$, CP applies the following learnable (non-linear) mapping 
\begin{equation}
    g_Q(\cdot;\theta_Q):
    \mathbb{C}^{NM}
    \rightarrow
    \mathbb{C}^{NM}
    \label{eq:quantizer_mapping}
\end{equation}
to $\mathbf{x}_i$, yielding $g_Q(\mathbf{x}_i;\theta_Q).$ This pre-processed signal will then go through the scalar uniform quantization $Q_\text{uni}$, i.e., (cf.~\ref{eq:quantized_signal})
\begin{equation}
    \tilde{\mathbf{q}}_i =  Q_\text{uni}(g_Q(\mathbf{x}_i;\theta_Q)).
\end{equation}

Upon reception of $\tilde{\mathbf{q}}_i^m$, the $m$-th RU will first apply the standard scalar uniform dequantization, i.e., $Q_{m,\text{uni}}^{-1}(\tilde{\mathbf{q}}_i^m)$. It then applies its own learnable (non-linear) mapping
\begin{equation}
    g_{Q,m}(\cdot;\theta_{Q,m}):\mathbb{C}^N \rightarrow \mathbb{C}^N
\end{equation}
to $Q_{m,\text{uni}}^{-1}(\tilde{\mathbf{q}}_i^m)$, yielding $g_{Q,m}(Q_{m,\text{uni}}^{-1}(\tilde{\mathbf{q}}_i^m);\theta_{Q,m})$. Such reconstructed signal will then be transmitted to the UEs, i.e., (cf.~\ref{eq:overall_mapping})
\begin{equation}
    \mathbf{y}_i = \sum_{m=1}^{M}  \mathbf{H}_{m} \cdot  g_{Q,m}(Q_{m,\text{uni}}^{-1}(\tilde{\mathbf{q}}_i^m);\theta_{Q,m}) + \mathbf{n}_i.
\end{equation}
The RU-side post-processing is designed such that the reconstructed
signal satisfies the per-RU transmit-power constraint, i.e., 
$\mathbb{E}[\|g_{Q,m}(Q_{m,\text{uni}}^{-1}(\tilde{\mathbf{q}}_i^m);
\theta_{Q,m})\|^2]\leq P_{\rm tx}$ for all $m$.

In order to find an adequate learnable mappings that allow minimal performance degradation due to the finite fronthaul capacity, we consider the following optimization problem
\begin{equation}
\label{eq:ns_quantizer_training_problem}
\begin{aligned}
    &\max_{\theta_Q, \{\theta_{Q,m}\}_{m=1}^M}
    \quad
    \mathbb{E}_{\mathbf{H}}\big[
    R_{\mathrm{q},\epsilon}(\mathbf{H}, Q, \{Q_m^{-1}\}_{m=1}^M)
    \big] \\
    &\text{s.t.} \,\,
    Q(\cdot)=Q_{\text{uni}}\circ g_Q(\cdot;\theta_Q), \\
    &\phantom{\text{s.t.}\,\,}
    Q_m^{-1}(\cdot)=g_{Q,m}(\cdot;\theta_{Q,m})\circ
    Q_{m,\text{uni}}^{-1}, \,\forall m=1,\ldots,M,
\end{aligned}
\end{equation}
where $R_{\mathrm{q},\epsilon}(\mathbf{H}, Q, \{Q_{m}^{-1}\}_{m=1}^M)\equiv R_\epsilon(T; P_{\mathbf{y}^\text{WMMSE}|\mathbf{s}})$  whose conditional distribution $P_{\mathbf{y}^\text{WMMSE}|\mathbf{s}}$ is given by (\ref{eq:quantized_signal})--(\ref{eq:overall_mapping}) that adopts a conventional WMMSE precoder (Algorithm~\ref{alg:wmmse_refinement}), initialized with a precoding matrix generated by randomly sampling each element independently from $\mathcal{CN}(0,1)$. We refer to Appendix~\ref{quantizer_design_eval_appendix} for further details on the practical approximation of $R_{\mathrm{q},\epsilon}(\mathbf{H}, Q, \{Q_{m}^{-1}\}_{m=1}^M)$ that considers Bussgang decomposition \cite{bussgang1952crosscorrelation} along with the worst case (in terms of the first-order asymptotics) uncorrelated additive noise theorem \cite{hassibi2003much}. In a manner similar to the previous subsection, we approximate the expectation in (\ref{eq:ns_quantizer_training_problem}) with the available channel realizations collected offline. 


\subsubsection{Time complexity}
The non-scalable AI-based quantizer does not require online quantizer design
time, since the quantization-dequantization functions in
\eqref{eq:ns_quantizer_training_problem} are
trained offline and remain unchanged during online transmission. Thus, the CP does not
design a new quantization function within each coherence block, hence the
quantizer design time in \eqref{eq:objective} can be set to $t_2=0$. In
addition, the RU-side dequantization functions are shared before deployment and
are not updated through fronthaul signaling during online transmission, making the 
dequantizer-update signaling time as  $t_3=0$. 

The remaining online computation is the per-symbol evaluation of the CP-side
mapping $g_Q(\cdot;\theta_Q)$, scalar uniform quantization, scalar uniform
dequantization, and the matched RU-side post-processing functions
$\{g_{Q,m}(\cdot;\theta_{Q,m})\}_{m=1}^{M}$. Let $t_{Q}$ denote the
hardware-specific wall-clock time to evaluate these operations for one
transmitted symbol vector. For instance, when the CP-side mapping and RU-side
post-processing functions are implemented by MLPs with $L_Q$ hidden layers each
with $d_Q$ hidden neurons, $t_{Q}$ is dictated by the corresponding
forward-pass computational complexity, which is of order
$\mathcal{O}(NMd_Q+L_Qd_Q^2)$ per transmitted symbol vector. Meanwhile, the scalar uniform
quantization and its matched scalar
dequantization, introduced in Appendix~\ref{quantizer_design_eval_appendix}, require only
$\mathcal{O}(NM)$ additional operations which is generally negligible.

\section{Test-Time Scalable AI-Based Modules}
\label{sec:test_time_scalable_ai_modules}

In this section, we extend Sec.~\ref{sec:non_scalable_ai_modules} so that the
AI-based modules can achieve test-time scalability for the targeted performance
metric. To start, we introduce the generic definition of test-time scalability.

\begin{definition}[Test-time scalability]
An AI module that yields a decision $D(t)$ using inference time $t$ is called
test-time scalable with respect to a metric $R$ if
\begin{equation}
    R(D(t')) \geq R(D(t)),
\end{equation}
for any $t'\geq t$, and the inequality is strict for at least one pair
$(t,t')\in\mathbb{R}_+^2$ with $t'>t$.
\label{tts_def}
\end{definition}

\subsection{Test-Time Scalable AI-Based Precoder Design}
\label{subsec:tts_ai_precoder}

In Sec.~\ref{subsec:non_scalable_ai_precoder}, the AI-assisted precoder design
uses a fixed inference time $t_1=t_{\rm init}+I_Wt_{\rm WMMSE}$ to generate a
single WMMSE initialization. In this subsection, we extend the AI-assisted
WMMSE algorithm so that it achieves test-time scalability with respect to the
finite-blocklength sum-rate $R_{{\rm p},\epsilon}$. The key idea is to train a
stochastic initialization rule that generates multiple candidate precoders and
then chooses the one that best achieves the target metric.

Therefore, we consider a stochastic initialization rule
$G_W(\cdot;\theta_W')$ whose inference time is nearly the same as that of the
deterministic initializer $g_W(\cdot;\theta_W)$ in
Sec.~\ref{subsec:non_scalable_ai_precoder}. Here, $\theta_W'$ denotes the
learnable parameter vector of the stochastic initializer, which includes the
parameters required to generate randomized initial precoders. Specifically, we
define
\begin{equation}
    G_W(\cdot;\theta_W'):
    \mathcal{H}
    \rightarrow
    \mathcal{P}_{\mathcal{W}},
    \label{eq:stochastic_precoder_initializer}
\end{equation}
where $\mathcal{P}_{\mathcal{W}}$ is the set of all probability
distributions on the feasible precoder space $\mathcal{W}$. In the sequel, we will describe how such a stochastic initialization rule $G_W(\cdot;\theta_W')$ can lead to the design of a test-time scalable AI-based precoder design. The key idea is to first generate multiple initial precoding matrices; run the WMMSE algorithm for each of the initializations; choose the one that would perform the best using an appropriate evaluation strategy.

Specifically, given some
channel matrix $\mathbf{H}$, the
$j$-th candidate precoder is obtained as
\begin{equation}
    \mathbf{W}_j
    =
    \mathrm{WMMSE}_{I_W}
    \big(
        \mathbf{W}_{0,j},\mathbf{H}
    \big),
    \quad
    \mathbf{W}_{0,j}\overset{\text{i.i.d.}}{\sim}G_W(\mathbf{H};\theta_W'),
    \label{eq:tts_precoder_candidate}
\end{equation}
for $j=1,\ldots, J_W(t)$, where $J_W(t)$ is the affordable number of candidate precoders under the computing time budget of $t$, defined as
\begin{equation}
    J_W(t)
    =
    \left\lfloor
    \frac{t}
    {t_{\rm init}+I_Wt_{\rm WMMSE}+t_{{\rm p},{\rm eval}}}
    \right\rfloor,
    \label{eq:num_precoder_candidates}
\end{equation}
where $t_{\rm init}$ is the time required to draw one initialization from
$G_W(\mathbf{H};\theta_W')$, $I_Wt_{\rm WMMSE}$ is the WMMSE refinement time,
and $t_{{\rm p},{\rm eval}}$ is the time required to evaluate
$R_{{\rm p},\epsilon}(\mathbf{H},\mathbf{W}_j)$ for the single candidate precoder $\mathbf{W}_j$. 

The proposed test-time scalable AI-based precoding algorithm $\mathcal{A}{\rm lg}_W(t;\mathbf{H})$ then chooses one of the candidate precoders as follows
\begin{equation}
\begin{array}{@{}l@{}}
\displaystyle
\mathcal{A}{\rm lg}_W(t;\mathbf{H})
=
\argmax\limits_{\mathbf{W}\in\{\mathbf{W}_j\}_{j=1}^{J_W(t)}}
R_{{\rm p},\epsilon}(\mathbf{H},\mathbf{W})
\\[1ex]
\displaystyle
\text{s.t.}\quad
\mathbf{W}_j
=
\mathrm{WMMSE}_{I_W}
\big(
    \mathbf{W}_{0,j},\mathbf{H}
\big),
\\[0.5ex]\quad\quad
\displaystyle
\mathbf{W}_{0,j}\overset{\rm i.i.d.}{\sim}G_W(\mathbf{H};\theta_W'),
\qquad j=1,\ldots,J_W(t).
\end{array}
\label{eq:tts_ai_wmmse_algorithm}
\end{equation}

We will now show that $\mathcal{A}{\rm lg}_W(t;\mathbf{H})$ satisfies Def. \ref{tts_def} in general. To this end, let us first fix the random seed for the sampling from $G_W(\mathbf{H};\theta_W')$ so that
\begin{equation}
    \{\mathbf{W}_{0,j}\}_{j=1}^{J_W(t)}
    \subseteq
    \{\mathbf{W}_{0,j}\}_{j=1}^{J_W(t')},
    \label{eq:precoder_candidate_set_inclusion}
\end{equation}
holds for any $t'\geq t$. Accordingly, we have
\begin{equation}
    R_{{\rm p},\epsilon}
    \big(
        \mathbf{H},
        \mathcal{A}{\rm lg}_W(t';\mathbf{H})
    \big)
    \geq
    R_{{\rm p},\epsilon}
    \big(
        \mathbf{H},
        \mathcal{A}{\rm lg}_W(t;\mathbf{H})
    \big),
    \label{tts_precoder_def}
\end{equation}
for any $t'\geq t$. Upon appropriate training (see Appendix \ref{appendix:tts_precoder_details}), \eqref{tts_precoder_def} generally admits strict inequality for some pairs $(t',t)$, which we empirically validate in Sec. \ref{experimental results} (see Fig. \ref{result_precoder_inference}). Thus, $  \mathcal{A}{\rm lg}_W(t;\mathbf{H})$ generally satisfies Def. \ref{tts_def} with respect to the metric $R_{{\rm p},\epsilon}(\mathbf{H},\cdot)$. We refer to Appendix~\ref{appendix:tts_precoder_details} for detailed implementation of the stochastic initializer $G_W(\cdot;\theta_W')$ inspired by the context-based meta-learning \cite{zintgraf2019fast}, reported to also be effective when dealing with time-varying channels \cite{park2020learning}.

 \begin{figure}[t]
         \centering
         \begin{subfigure}[t]{0.6\textwidth}
             \centering
             \includegraphics[width=\textwidth]{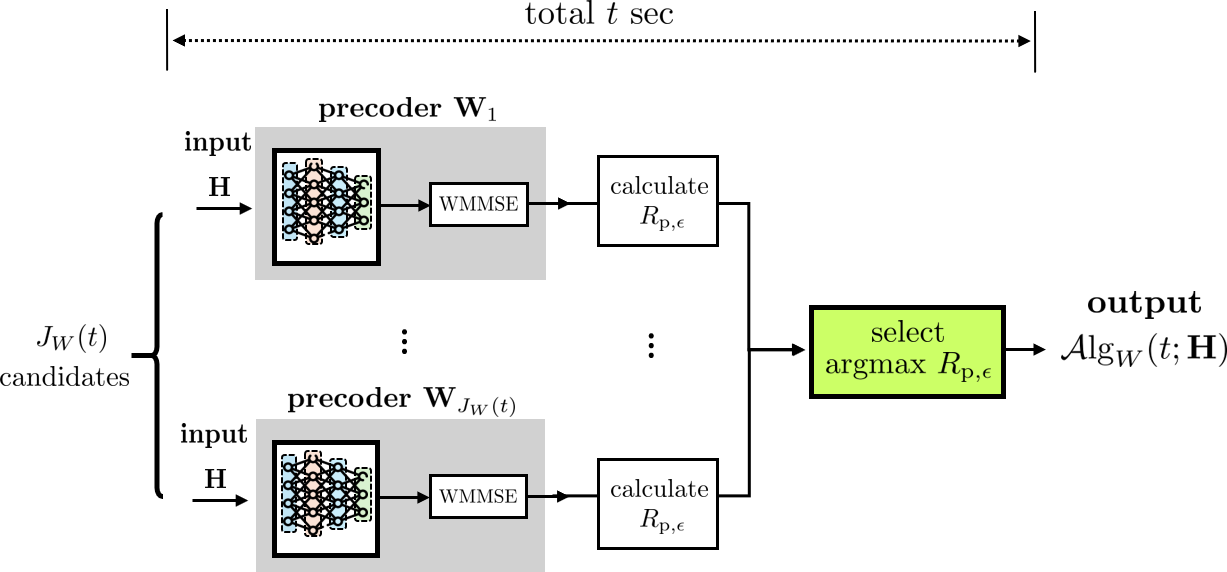}
             \caption{}
             \label{AI_precoder}
         \end{subfigure}\\ \vspace{5pt}
         \begin{subfigure}[t]{0.6\textwidth}
             \centering
             \includegraphics[width=\textwidth]{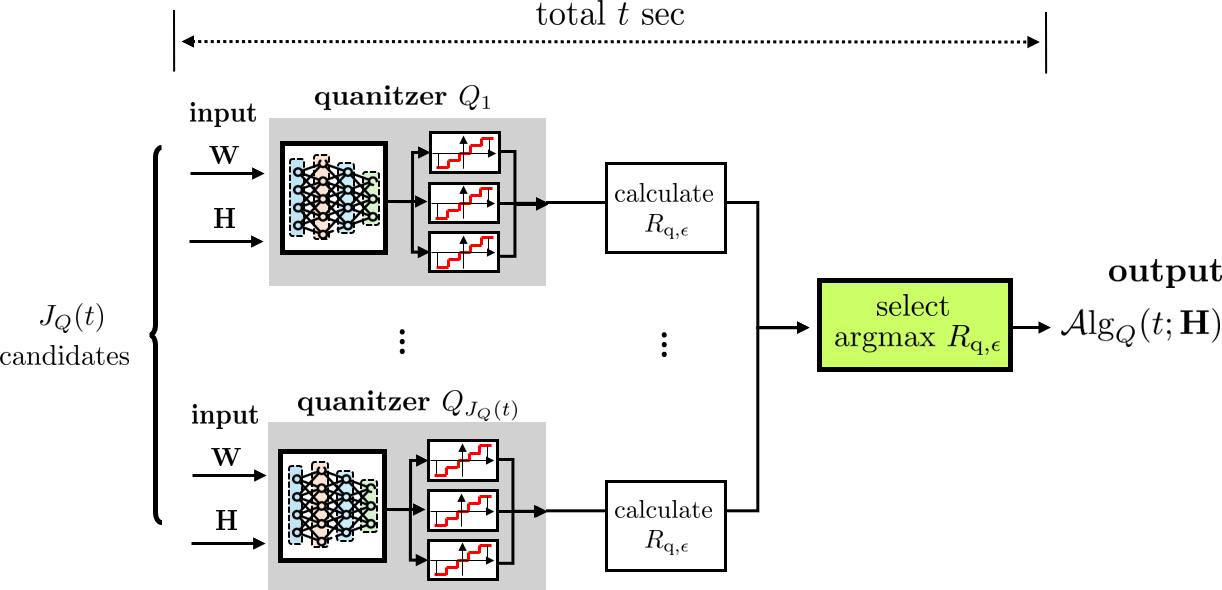}
            \caption{}
             \label{AI_quantizer}
         \end{subfigure}\\
            \caption{Test-time scalable AI-based precoder and fronthaul quantizer architecture. (a) The scalable precoder generates multiple initial precoding matrices from the stochastic initialization rule $G_W(\mathbf{H};\theta_W')$, refines each initialization using the WMMSE algorithm, and selects the candidate precoder that maximizes $R_{{\rm p},\epsilon}$. (b) The scalable fronthaul quantizer generates multiple matched quantization-dequantization function tuples from the stochastic parameter-generation rules $G_Q(\theta_Q')$ and $\{G_{Q,m}(\theta_{Q,m}')\}_{m=1}^{M}$, evaluates the candidates using $R_{{\rm q},\epsilon}$, and selects the best tuple.}
            \label{test_time_scalable_model}
    \end{figure}

\subsection{Test-Time Scalable AI-Based Quantizer Design}
\label{subsec:tts_ai_quantizer}

In Sec.~\ref{subsec:non_scalable_ai_quantizer}, the AI-assisted quantizer
design uses fixed quantization-dequantization functions learned offline. In this
subsection, we extend the AI-assisted deep task-based quantizer so that it
achieves test-time scalability with respect to the finite-blocklength sum-rate $R_{{\rm q},\epsilon}$. The key idea is to prepare multiple candidate pairs of CP-side pre-processing function and RU-side post-processing functions, then choose the one that best achieves the target metric at each coherence block.

To this end, we generate multiple candidate quantization-dequantization function
tuples by randomizing the parameters of the CP-side mapping and the RU-side
post-processing functions. Let
$\Theta_Q\subseteq\mathbb{R}^{d_Q}$ and
$\Theta_{Q,m}\subseteq\mathbb{R}^{d_{Q,m}}$ denote the parameter spaces of CP-side pre-processing function $g_Q(\cdot;\theta_Q)$ and $m$-th RU-side post-processing function $g_{Q,m}(\cdot;\theta_{Q,m})$, respectively. We
define
\begin{equation}
    G_Q:
    \mathbb{R}^{d_Q'}
    \rightarrow
    \mathcal{P}_{\Theta_Q},
    G_{Q,m}:
    \mathbb{R}^{d_{Q,m}'}
    \rightarrow
    \mathcal{P}_{\Theta_{Q,m}},
    \label{eq:stochastic_quantizer_generators}
\end{equation}
for $m=1,\ldots,M$, where $\mathcal{P}_{\Theta_Q}$ and $\mathcal{P}_{\Theta_{Q,m}}$ are the sets of
all probability distributions on $\Theta_Q$ and $\Theta_{Q,m}$, respectively.
The vectors $\theta_Q'\in\mathbb{R}^{d_Q'}$ and
$\theta_{Q,m}'\in\mathbb{R}^{d_{Q,m}'}$ are the learnable parameters of these
stochastic parameter-generation rules. We refer to Appendix~\ref{appendix:tts_quantizer_details} for further details on the actual implementation of the stochastic mapping via random noise injection, inspired by aleatoric uncertainty modeling in the model space that has shown remarkable performance in safety-critical scenarios.

In
the sequel, we describe how these stochastic parameter-generation rules lead to
the design of a test-time scalable AI-based quantizer. We first generate
multiple matched parameter sets; construct a quantization-dequantization tuple
from each parameter set; evaluate each tuple using $R_{{\rm q},\epsilon}$; and
choose the one that performs the best.

Given some channel $\mathbf{H}$, the $j$-th
candidate quantization-dequantization tuple is obtained as
\begin{equation}
\begin{aligned}
    &
    Q_j(\cdot)
    =
     Q_{\rm uni}\circ g_Q(\cdot;\theta_{Q,j}),
    \\
    &
    Q_{j,m}^{-1}(\cdot)
    =
    g_{Q,m}(\cdot;\theta_{Q,m,j})\circ
     Q_{m,{\rm uni}}^{-1}(\cdot),
    \\
    &
    \theta_{Q,j}
    \overset{\rm i.i.d.}{\sim}
    G_Q(\theta_Q'),
    \qquad
    \theta_{Q,m,j}
    \overset{\rm i.i.d.}{\sim}
    G_{Q,m}(\theta_{Q,m}'),
\end{aligned}
\label{eq:tts_quantizer_candidate_tuple}
\end{equation}
for $m=1,\ldots,M$ and $j=1,\ldots,J_Q(t)$, where $J_Q(t)$ is the affordable number of candidate tuples under the computing time budget $t$, defined as
\begin{equation}
    J_Q(t)
    =
    \left\lfloor
    \frac{t}
    {t_{\rm init}+t_{{\rm q},{\rm eval}}}
    \right\rfloor,
    \label{eq:num_quantizer_candidates}
\end{equation}
where $t_{\rm init}$ is the time required to generate one matched parameter set
through $G_Q$ and $\{G_{Q,m}\}_{m=1}^{M}$ and construct the quantization-dequantization candidate tuple, and $t_{{\rm q},{\rm eval}}$ is the time required to evaluate
$R_{{\rm q},\epsilon}$ for the single candidate
$(Q_j,\{Q_{j,m}^{-1}\}_{m=1}^{M})$. 

The proposed test-time scalable AI-based quantization algorithm $\mathcal{A}{\rm lg}_Q(t;\mathbf{H})$ then chooses one of the candidate tuples as follows
\begin{equation}
\begin{aligned}
&
\mathcal{A}{\rm lg}_Q(t;\mathbf{H})
=
\\
&
\begin{array}{@{}l@{}}
\displaystyle
\argmax\limits_{
\begin{subarray}{c}
(Q,\{Q_m^{-1}\}_{m=1}^{M}) \in \\
\{(Q_j,\{Q_{j,m}^{-1}\}_{m=1}^{M})\}_{j=1}^{J_Q(t)}
\end{subarray}
}
R_{{\rm q},\epsilon}
\left(
    \mathbf{H},
    Q,
    \{Q_m^{-1}\}_{m=1}^{M}
\right)
\\[1ex]
\displaystyle
\text{s.t.}\quad
Q_j(\cdot)
=
Q_{\rm uni}\circ g_Q(\cdot;\theta_{Q,j}),\\
\qquad\,\,\,
Q_{j,m}^{-1}(\cdot)=g_{Q,m}(\cdot;\theta_{Q,m,j})\circ
Q_{m,\text{uni}}^{-1}(\cdot),\\
\qquad\,\,\, \theta_{Q,j}\overset{\text{i.i.d.}}{\sim}G_Q(\theta_Q'), \quad \theta_{Q,m,j}\overset{\text{i.i.d.}}{\sim}G_{Q,m}(\theta_{Q,m}'),
\end{array}
\label{eq:tts_ai_quantizer_algorithm}
\end{aligned}
\end{equation}
for $j=1,\ldots,J_Q(t)$. We will now show that $\mathcal{A}{\rm lg}_Q(t;\mathbf{H})$
satisfies Def.~\ref{tts_def} in general. To this end, let us first fix the random seed for the sampling from $G_Q(\theta_Q')$ and $\{G_{Q,m}(\theta_{Q,m}')\}_{m=1}^{M}$ so that \begin{equation} \left\{ (\theta_{Q,j},\{\theta_{Q,m,j}\}_{m=1}^{M}) \right\}_{j=1}^{J_Q(t)} \subseteq \left\{ (\theta_{Q,j},\{\theta_{Q,m,j}\}_{m=1}^{M}) \right\}_{j=1}^{J_Q(t')}, \label{eq:quantizer_candidate_set_inclusion} \end{equation} for any $t'\geq t$. Accordingly, we have
\begin{equation}
\begin{aligned}
    R_{{\rm q},\epsilon}
    &\left(
        \mathbf{H},
        \mathcal{A}{\rm lg}_Q(t';\mathbf{H})
    \right)\geq 
    R_{{\rm q},\epsilon}
    \left(
        \mathbf{H},
        \mathcal{A}{\rm lg}_Q(t;\mathbf{H})
    \right),
\end{aligned}
    \label{tts_quantizer_def}
\end{equation}
for any $t'\geq t$. Upon appropriate training
(see Appendix~\ref{appendix:tts_quantizer_details}), \eqref{tts_quantizer_def} generally
admits strict inequality for some pairs $(t',t)$, which is empirically
validated in Sec.~\ref{experimental results} (see Fig. \ref{result_quantizer_inference}). Hence,
$\mathcal{A}{\rm lg}_Q(t;\mathbf{H})$ also generally satisfies
Def.~\ref{tts_def} with respect to the metric
$R_{{\rm q},\epsilon}(\mathbf{H},\cdot)$.


\section{Test-Time Scalable AI-Based Cell-Free MIMO}
\label{sec:test_time_scalable_cell_free_mimo}

The (separately designed) scalable modules in Sec.~\ref{sec:test_time_scalable_ai_modules} generate
candidate outputs for the precoder $\mathbf W$ and the
quantization-dequantization functions $(Q,\{Q_m^{-1}\}_{m=1}^{M})$,
respectively. However, the received signal in \eqref{eq:overall_mapping}
depends on the joint combination of $\mathbf W$, $Q$, and $\{Q_m^{-1}\}_{m=1}^{M}$.
Hence, independently selecting the precoder and the
quantization-dequantization functions need not be necessarily optimal for the
end-to-end downlink performance. In this section, we propose the joint selection rule for $\mathbf W$, $Q$, and
$\{Q_m^{-1}\}_{m=1}^{M}$ whose inference times are optimally allocated based on the long-term statistics of the time-varying channels.  


\subsection{Joint Selection for Test-Time Scalability}
\label{subsec:joint_selection}

Let $t_1$ and $t_2$ denote the inference times allocated to the scalable precoder and quantizer, respectively. The signaling time $t_3$ considered in Section~\ref{subsec:goal} is omitted
here since the selected candidate index incurs negligible fronthaul signaling
overhead. For a given channel matrix $\mathbf H$ and inference time $t_1$, the scalable precoder in Sec.~\ref{subsec:tts_ai_precoder} generates the candidate precoders $\{\mathbf W_j\}_{j=1}^{J_W(t_1)}$ according to \eqref{eq:tts_precoder_candidate}. Similarly, during inference time $t_2$, the scalable quantizer in Sec.~\ref{subsec:tts_ai_quantizer} generates the candidate quantization-dequantization functions $ \{ (Q_\ell,\{Q_{\ell,m}^{-1}\}_{m=1}^{M}) \}_{\ell=1}^{J_Q(t_2)}$ according to \eqref{eq:tts_quantizer_candidate_tuple}. Since the received signal in \eqref{eq:overall_mapping} depends on the combination of $\mathbf W$, $Q$, and $\{Q_m^{-1}\}_{m=1}^{M}$, these candidates should be selected based on their joint effect on the end-to-end system performance.


As a starting point, let $R_{\epsilon}(\mathbf H,\mathbf W,Q,\{Q_m^{-1}\}_{m=1}^{M}) \equiv R_\epsilon(T; P_{\mathbf{y}|\mathbf{s}})$  where the conditional distribution $P_{\mathbf{y}|\mathbf{s}}$ is given by (\ref{eq:quantized_signal})--(\ref{eq:overall_mapping}) that uses $\mathbf{W},Q,$ and $\{Q_m^{-1}\}_{m=1}^{M}$. Now, given some arbitrary pair of $(t_1,t_2,\mathbf H)$, the proposed joint selection algorithm $\mathcal{A}{\rm lg}_{\rm AI}(t_1,t_2;\mathbf H)$ can be written as 
\begin{equation}
\begin{aligned}
&
\mathcal{A}{\rm lg}_{\rm AI}(t_1,t_2;\mathbf H)
=
\\
&
\begin{array}{@{}l@{\quad}l@{}}
\multicolumn{2}{@{}l@{}}{
\displaystyle
\argmax\limits_{
\substack{
\mathbf W\in\{\mathbf W_j\}_{j=1}^{J_W(t_1)},\\
(Q,\{Q_m^{-1}\}_{m=1}^{M})
\in
\{(Q_\ell,\{Q_{\ell,m}^{-1}\}_{m=1}^{M})\}_{\ell=1}^{J_Q(t_2)}
}}
\hspace{-1.5cm}R_{\epsilon}
\left(
    \mathbf H,
    \mathbf W,
    Q,
    \{Q_m^{-1}\}_{m=1}^{M}
\right)
}
\\[1ex]
\displaystyle
\hspace{1.2cm}\mathrm{s.t.}
&
\displaystyle
\mathbf W_j
=
\mathrm{WMMSE}_{I_W}
\left(
    \mathbf W_{0,j},
    \mathbf H
\right),
\\[1ex]
&
\displaystyle
Q_\ell(\cdot)
=
 Q_{\rm uni}\circ g_Q(\cdot;\theta_{Q,\ell}),
\\[1ex]
&
\displaystyle
Q_{\ell,m}^{-1}(\cdot)
=
g_{Q,m}(\cdot;\theta_{Q,m,\ell})\circ
 Q_{m,{\rm uni}}^{-1}(\cdot),
\\[1ex]
&
\begin{array}{@{}l@{\quad}l@{}}
\displaystyle
\mathrm{s.t.}
&
\displaystyle
\mathbf W_{0,j}
\overset{\rm i.i.d.}{\sim}
G_W(\mathbf H;\theta_W'),
\\[1ex]
&
\displaystyle
\theta_{Q,\ell}
\overset{\rm i.i.d.}{\sim}
G_Q(\theta_Q'),
\\[1ex]
&
\displaystyle
\theta_{Q,m,\ell}
\overset{\rm i.i.d.}{\sim}
G_{Q,m}(\theta_{Q,m}'),
\end{array}
\end{array}
\end{aligned}
\label{eq:joint_ai_output}
\end{equation}
where $J_W(t_1)$ and $J_Q(t_2)$ are defined in \eqref{eq:num_precoder_candidates} and \eqref{eq:num_quantizer_candidates}, respectively. The rule in \eqref{eq:joint_ai_output} performs a joint rate-based selection
over the precoder candidates and the quantization-dequantization candidates
generated by the separately designed scalable modules.

With the fixed random seed sequences used in
Sec.~\ref{subsec:tts_ai_precoder} and~\ref{subsec:tts_ai_quantizer},
increasing $t_1$ or $t_2$ only appends additional candidates to the
corresponding candidate list. Hence, for any $t_1'\geq t_1$ and
$t_2'\geq t_2$, the search space in \eqref{eq:joint_ai_output} never shrinks, and
therefore
\begin{equation}
R_{\epsilon}
\left(
    \mathbf H,
    \mathcal{A}{\rm lg}_{\rm AI}(t_1,t_2;\mathbf H)
\right)
\leq
R_{\epsilon}
\left(
    \mathbf H,
    \mathcal{A}{\rm lg}_{\rm AI}(t_1',t_2';\mathbf H)
\right).
\label{eq:joint_tts_monotonicity}
\end{equation} 
Upon appropriate training, the
inequality in \eqref{eq:joint_tts_monotonicity} generally becomes strict for
some pairs $(t_1',t_2')$ and $(t_1,t_2)$, which is empirically validated in Sec.~\ref{subsec:joint_time_allocation_results_1} (see Fig. \ref{result_joint_scalable}). Hence,  $\mathcal{A}{\rm lg}_{\rm AI}(t_1,t_2;\mathbf H)$ is jointly test-time scalable
with respect to $R_{\epsilon}$ in general.




\subsection{Optimal Resource Allocation}
\label{subsec:optimal_resource_allocation}
The joint selection algorithm in \eqref{eq:joint_ai_output} allows us to maximally utilize the (separately) designed AI modules for any given pair of $(t_1,t_2)$ -- the allocated inference times for the test-time scalable AI modules. This can be treated as an approximate solution of the objective in Sec.~\ref{subsec:goal} which aims at finding the best pair of precoding and quantization-dequantization functions given the fixed computing/signaling time allocations. Having such an approximate solution at hand, we are now ready to formulate the inference time allocation problem as briefly mentioned in Sec.~\ref{subsec:goal}:
\begin{align}
    \max_{t_1, t_2 \in [0,T]}& \ \ 
    \mathbb{E}_{\mathbf{H}_{\rm prev}, \mathbf{H}}
    \big[
    T_\text{prep}\cdot
    R_{\epsilon}(T_\text{prep};P_{\mathbf{y}^\text{outdated}|\mathbf{s}})
    \nonumber\\
    &\quad
    +  (T-T_\text{prep}) \cdot
    R_{\epsilon}(T-T_\text{prep};P_{\mathbf{y}|\mathbf{s}})
    \big]
    \nonumber\\
    \mathrm{s.t.}\quad 
    &T_\text{prep} = t_1 + t_2\leq T,
    \nonumber\\
    &
    \{ \mathbf{W}, Q, \{Q_m^{-1}\}_{m=1}^M\}
    =
    \mathcal{A}{\rm lg}_{\text{AI}}(t_1, t_2;\mathbf{H}), \nonumber\\&
     \{ \mathbf{W}_\text{prev}, Q_\text{prev}, \{Q_{\text{prev},m}^{-1}\}_{m=1}^M\}
    \hspace{-2pt}=\hspace{-2pt}
    \mathcal{A}{\rm lg}_{\text{AI}}(t_1, t_2;\mathbf{H}_\text{prev}), \label{eq:tts_resource_allocation}
\end{align}
In \eqref{eq:tts_resource_allocation}, the conditional
distribution $P_{\mathbf{y}|\mathbf{s}}$ is given by
\eqref{eq:quantized_signal}--\eqref{eq:overall_mapping} with the current
channel $\mathbf H$ and the selected current-block functions
$\{ \mathbf{W}, Q(\cdot), \{Q_m^{-1}\}_{m=1}^M\}$. The outdated conditional
distribution $P_{\mathbf{y}^\text{outdated}|\mathbf{s}}$ is also given by
\eqref{eq:quantized_signal}--\eqref{eq:overall_mapping}, but with the current
channel $\mathbf H$ and the previous-block (outdated) functions
$\{ \mathbf{W}_{\rm prev}, Q_{\rm prev}(\cdot),
\{Q_{m,\rm prev}^{-1}\}_{m=1}^M\}
=
\mathcal{A}{\rm lg}_{\text{AI}}(t_1, t_2;\mathbf{H}_{\rm prev})$.
The expectation is taken over the temporal channel variation across
consecutive coherence blocks \cite{Emil2017Temporal_correlation}, i.e., the joint distribution of
$(\mathbf H_{\rm prev},\mathbf H)$. In practice, the expectation in
\eqref{eq:tts_resource_allocation}  is approximated through empirical averaging. We solve  \eqref{eq:tts_resource_allocation} by grid search \cite{Boshkovska2017gridsearch} over
sufficiently fine-grained grids
$[0,\Delta_t,2\Delta_t,\ldots,G\Delta_t]\times
[0,\Delta_t,2\Delta_t,\ldots,G\Delta_t]$, where $\Delta_t$ denotes the inference time
grid resolution and $G=\lfloor T/\Delta_t\rfloor$ is the corresponding grid
size. We refer to Appendix~\ref{appendix:tts_precoder_details} and \ref{appendix:tts_quantizer_details} for further details.


    \section{Experimental Results}
    \label{experimental results}

\subsection{Simulation Setting}
\label{subsec:simulation_setting}

    
We first summarize the simulation setup used to evaluate the
proposed test-time scalable AI-RAN framework. For channel generation, we adopt
the realistic outdoor ray-tracing dataset provided by Deep-MIMO
\cite{alkhateeb2019deepmimo}. The candidate UE locations
provided by the considered Deep-MIMO scenario are then divided into disjoint training
and test sets. During training, the UEs are sampled from the
training set, whereas during testing, they are sampled from the held-out test
set under the same RU deployment and propagation environment. 

To capture temporal channel variations due to user mobility, we model the
channel matrix $\mathbf H$ by applying a first-order channel aging process to
the small-scale fading vectors that determine the RU-UE channel components.
Specifically, following \cite{Emil2017Temporal_correlation}, for the small-scale fading vector between UE $k$ and RU $m$, we
set 
\begin{equation}
    \boldsymbol{\alpha}_{k,m}
    =
    \rho \boldsymbol{\alpha}_{k,m}^{\rm prev}
    +
    \sqrt{1-\rho^2}\mathbf{z}_{k,m},
    \label{eq:temporal_correlation}
\end{equation}
where $\boldsymbol{\alpha}_{k,m}\in\mathbb{C}^{P}$ collects the small-scale
fading coefficients of the $P$ propagation paths between UE $k$ and RU $m$,
$\boldsymbol{\alpha}_{k,m}^{\rm prev}$ denotes its value in the previous
coherence block, and
$\mathbf{z}_{k,m}\sim\mathcal{CN}(\mathbf{0},\mathbf{I}_{P})$ is a Gaussian
innovation vector. Following the widely used Jakes model
\cite{jakes1974microwave}, the temporal correlation coefficient is set as
$\rho=J_0(2\pi f_D T)$, where $J_0(\cdot)$ denotes the zeroth-order Bessel
function of the first kind and $f_D$ is the maximum Doppler frequency. We set $f_D=78$ Hz.



Throughout the experiments, we set the carrier frequency $f_c=28$ GHz, bandwidth
$B=50$ MHz, $M=4$ RUs, each equipped with $N=64$ antenna elements, and
$K=4$ single-antenna UEs by default. The per-RU transmit power and fronthaul
capacity are set to $P_{\rm tx}=23$ dBm and $C_F=100$ Gbps, respectively. The coherence block duration is set to $T=50$ ms, and we consider $N_{\rm cb}=20$ coherence blocks when evaluating impacts due to the temporal channel variations  (\ref{eq:temporal_correlation}). The tolerable decoding error probability is set to $\epsilon=10^{-3}$.


The precoder initializer $g_W(\cdot;\theta_W)$ is implemented as
a residual MLP \cite{li2025AI_RAN_parameter} with depth $4$ and hidden
dimension $384$. For the quantization module, the CP-side mapping
$g_Q(\cdot;\theta_Q)$ and the RU-side post-processing functions
$\{g_{Q,m}(\cdot;\theta_{Q,m})\}_{m=1}^{M}$ in the quantization-dequantization module are implemented
as MLPs with depth $3$ and hidden dimension $256$, followed by $b$-bit uniform
scalar quantizers. The learning rates are set to $10^{-3}$ and $7\times10^{-4}$ for the
precoder and quantizer networks, respectively. For
both modules, the AdamW optimizer is used, and the dropout rate is set to
$0.1$. All the inference (execution) times are measured on an NVIDIA RTX 3090 GPU. We assume that the
hyperparameters required for candidate generation, including quantization bit resolutions, clipping rules, and quantization-dequantization architectures,
are shared offline between the CP and RUs before online deployment.

\subsection{Test-Time Scalability of the Precoding Module}
\label{subsec:precoder_scalability_results}

\begin{figure}[t]
    \centering
    \includegraphics[width=.65\columnwidth]{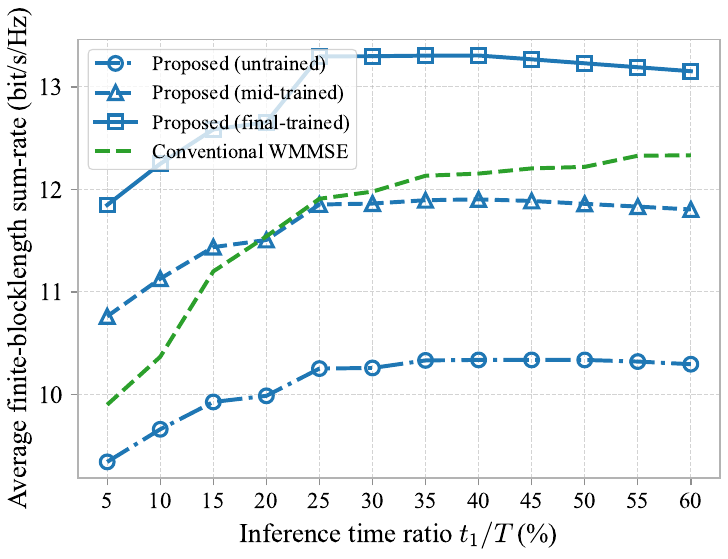}
    \caption{Average finite-blocklength sum-rate versus the precoder
    inference time ratio $t_1/T$ with $M=4$ RUs, $N=64$ antennas per RU, and
    $K=4$ single-antenna UEs under infinite fronthaul capacity.}
    \label{result_precoder_inference}
\end{figure}

First, we consider the precoder-only scenario under infinite fronthaul
capacity, which provides an ideal setting to examine the impact of the proposed
test-time scalable precoder design. Fig.~\ref{result_precoder_inference} shows
the average finite-blocklength sum-rate versus the precoder inference time
ratio $t_1/T$. In the untrained case, the stochastic initializer
$G_W(\mathbf H;\theta_W')$ is not optimized, and thus generates non-informative 
initial precoders, resulting in the lowest sum-rate. As training proceeds, the
performance improves, implying that the stochastic initializer finds an appropriate candidate set that contains a good precoding matrix. 

For the conventional WMMSE benchmark, the same inference time $t_1$ is used
for a single WMMSE run, so a longer inference time allows more WMMSE
refinement iterations. In contrast, for the proposed scalable precoder, the
number of WMMSE iterations per candidate is fixed to $I_W$ as in
\eqref{eq:tts_precoder_candidate}, while increasing $t_1$ increases the number
of candidate precoders $J_W(t_1)$ according to
\eqref{eq:num_precoder_candidates}. Therefore, the proposed method first
improves as $t_1$ increases, since more candidate precoders can be explored,
and then decreases when the reduced transmission duration within the coherence
block becomes dominant. This shows the tradeoff between precoder quality and
available data transmission time. The generalization capability of the trained stochastic initializer can be found in 
Appendix~\ref{appendix_outdated}.
    
\subsection{Test-Time Scalability of the Quantization Module}
\label{subsec:quantizer_scalability_results}

\begin{figure}[t]
    \centering
    \includegraphics[width=.65\columnwidth]{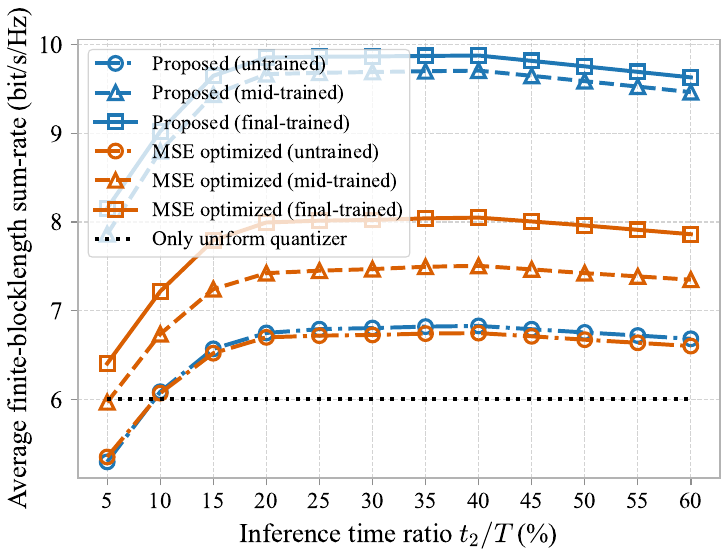}
    \caption{Average finite-blocklength sum-rate versus the quantizer
    inference time ratio $t_2/T$ with $M=4$ RUs, $N=64$ antennas per RU, and
    $K=4$ single-antenna UEs under fronthaul capacity $C_F=100$ Gbps.}
    \label{result_quantizer_inference}
\end{figure}

Next, we fix the precoder as the conventional WMMSE solution and consider the
quantizer-only scenario, where the proposed test-time scalable design is
applied only to the fronthaul quantizer. Fig.~\ref{result_quantizer_inference}
shows the average finite-blocklength sum-rate versus the quantizer
inference time ratio $t_2/T$. In the proposed method, candidate
quantization-dequantization tuples are generated as in
\eqref{eq:tts_quantizer_candidate_tuple} and selected according to the
finite-blocklength sum-rate metric in \eqref{eq:tts_ai_quantizer_algorithm}. For comparison, we also consider a mean-squared error (MSE)-based candidate selection rule that uses the same candidate set but selects the candidate minimizing the Euclidean distance between the original and the dequantized ones.


As shown in Fig.~\ref{result_quantizer_inference}, the proposed scalable
quantizer consistently outperforms the conventional uniform quantizer. As
$t_2$ increases, the number of candidate tuples $J_Q(t_2)$ in
\eqref{eq:num_quantizer_candidates} increases, and the sum-rate initially
improves. Beyond a certain point, however, the reduced transmission duration
within the coherence block becomes dominant, leading to a decrease in the sum
rate. The proposed rate-driven selection also outperforms the MSE-based
benchmark, showing the benefit of
selecting the quantizer according to the finite-blocklength sum-rate metric
rather than reconstruction accuracy alone. The generalization capability of the stochastic quantizer inference can be found 
in Appendix~\ref{appendix_outdated}.

\begin{figure}[t]
    \centering
    \includegraphics[width=.65\columnwidth]{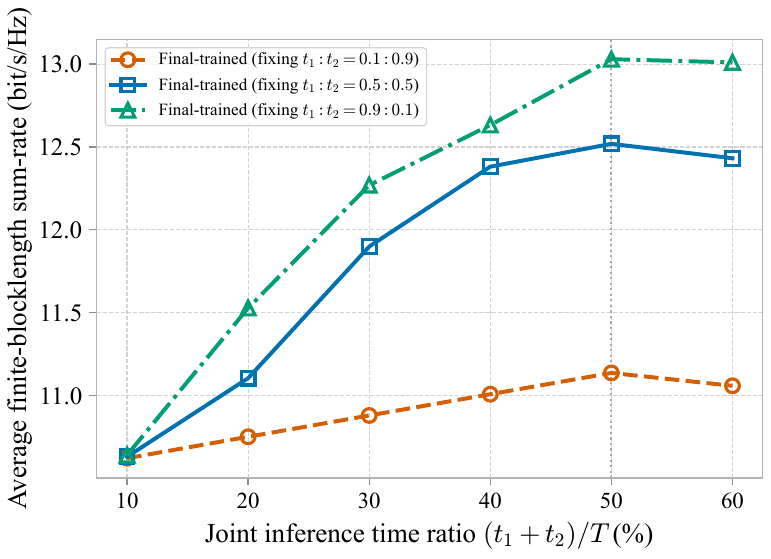}
    \caption{Average finite-blocklength sum-rate versus the  joint inference time ratio $(t_1+t_2)/T$ for different allocation ratios between $t_1$ and
$t_2$ with $M=4$ RUs, $N=64$ antennas per RU, and
    $K=4$ single-antenna UEs under fronthaul capacity $C_F=100$ Gbps.}
    \label{result_joint_scalable}
\end{figure}

\begin{figure}[t]
    \centering
    \includegraphics[width=.65\columnwidth]{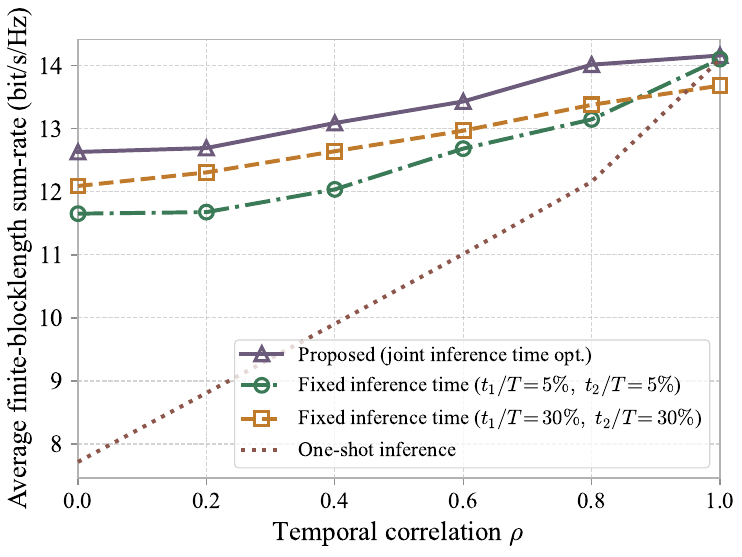}
    \caption{Average finite-blocklength sum-rate versus the temporal
    correlation coefficient $\rho$ with $M=4$ RUs, $N=64$ antennas per RU, and
    $K=4$ single-antenna UEs under fronthaul capacity $C_F=100$ Gbps.}
    \label{result_joint_inference}
\end{figure}

\subsection{Joint Test-Time Scalability}
\label{subsec:joint_time_allocation_results_1}
We now consider the joint precoding and quantization scenario by combining the separately designed  test-time scalable precoder in
Sec.~\ref{subsec:precoder_scalability_results} and test-time scalable quantizer in Sec.~\ref{subsec:quantizer_scalability_results}. As shown in Fig.~\ref{result_joint_scalable}, the average finite-blocklength
sum-rate increases as the normalized joint inference time ratio $(t_1+t_2)/T$
increases, even with the separately designed test-time scalable modules. However, the performance depends on the allocation ratio between
$t_1$ and $t_2$, showing that different splits of the same total test time can
lead to different sum-rate performance. This observation motivates the joint
optimization of $t_1$ and $t_2$, as demonstrated in the next subsection.

\subsection{Importance of Optimizing Inference Times $(t_1,t_2)$}
\label{subsec:joint_time_allocation_results}
Fig.~\ref{result_joint_inference} shows the average finite-blocklength sum
rate versus the temporal correlation coefficient $\rho$ in
\eqref{eq:temporal_correlation}. The proposed method selects the precoder and
quantization-dequantization functions jointly according to
\eqref{eq:joint_ai_output}, with the inference time allocation determined by
\eqref{eq:tts_resource_allocation}. The \emph{one-shot} benchmark reuses the precoder and quantization-dequantization
functions selected from the CSI of the first coherence block for all subsequent
coherence blocks. When $\rho=1$, the channel remains unchanged, and this
one-shot design remains valid across coherence blocks. In this regime, the
fixed-budget scheme with a small inference time allocation, e.g.,
$t_1/T=t_2/T=5\%$, also outperforms the scheme with a larger allocation, e.g.,
$t_1/T=t_2/T=30\%$, because additional computation provides limited gain while
reducing the remaining data transmission duration.

As $\rho$ decreases, the channel varies more rapidly, and the one-shot design
becomes increasingly outdated. In this case, allocating more inference time
can be beneficial because the selected precoder and quantization-dequantization
functions can better adapt to the current coherence block. Therefore, the best
choice of $(t_1,t_2)$ depends on the temporal channel correlation. The proposed
allocation in \eqref{eq:tts_resource_allocation} balances the quality of the
selected precoder and quantization-dequantization functions with the remaining
data transmission duration, and achieves the highest sum-rate irrespective of the 
values of $\rho$.

\begin{figure}[t]
    \centering
    \includegraphics[width=.65\columnwidth]{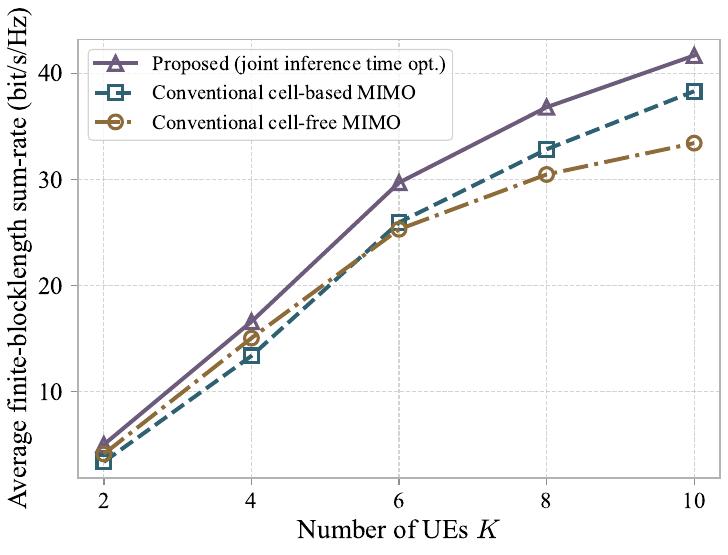}
    \caption{Average finite-blocklength sum-rate versus the number of UEs $K$
    with $M=4$ RUs and $N=64$ antennas per RU under fronthaul capacity
    $C_F=100$ Gbps.}
    \label{result_cell_free}
\end{figure}

\subsection{Cell-Free MIMO vs. Cell-Based MIMO}
Lastly, we compare the proposed framework with (\emph{i}) conventional, non-AI-based, cell-free MIMO and (\emph{ii}) conventional, non-AI-based, cell-based MIMO. That is, each of the benchmarks use conventional WMMSE precoding with random initialization via standard Gaussian (see (\ref{eq:ns_quantizer_training_problem})) followed by uniform fronthaul quantization (without any pre-processing). We set $t_1/T=15\%$ and $t_2=0$ accordingly.

Fig.~\ref{result_cell_free} shows the average finite-blocklength sum
rate versus the number of served UEs. In the regime of moderate number of UEs $(K < 6)$ conventional cell-free MIMO outperforms conventional cell-based MIMO thanks to its increased degree of freedom (DoF), though it is this increased DoF that makes the precoding and quantization in cell-free MIMO a more complicated task \cite{pellaco2020wmmse_deepunfolding}, leading to even worse performance than cell-based MIMO under the fixed inference time budget $t_1/T=15\%$. In contrast, the proposed, test-time scalable-AI-based cell-free MIMO consistently outperforms cell-based MIMO. This highlights again the importance of the proposed test-time scalable AI-RAN in cell-free MIMO systems.



\section{Conclusion}
\label{Conclusion}
In this paper, we have proposed a generic framework that enables maximal utilization of separately designed test-time scalable AI modules in cell-free MIMO systems. We first focused on extending conventional, deterministic, precoding AI modules and quantization-dequantization AI modules to test-time scalable, stochastic, AI modules. Then, a joint performance metric inspired by finite-blocklength information theory has been considered to ensure the joint test-time scalability of separately designed AI modules. Finally, by noting that the optimal allocation of inference times for the AI modules should differ depending on the long-term statistics of the underlying wireless channels, an optimal resource allocation problem has been considered; solved via a grid search. Experimental results on realistic cell-free MIMO settings confirmed the importance of inference time optimization in test-time scalable AI-RAN-enabled cell-free MIMO systems. Future work may consider extending the proposed framework to a broader class of test-time scalable modules involving sequential decision-making  \cite{ellis2025theory}, together with hardware-wise scheduling strategies \cite{zhou2026timely}.


\appendix

\subsection{Evaluation Details of AI-based Precoder}
\label{precoder_eval_appendix}

Here, $R_{{\rm p},\epsilon}(\mathbf{H},\mathbf{W})$ is the finite-blocklength
sum-rate achieved by the precoder $\mathbf{W}$ for target decoding error
probability $\epsilon$. Unless otherwise specified, $n=L$, where
$L=\lfloor T/T_s\rfloor$ is the number of symbol vectors within one coherence
block defined in Sec.~II. With i.i.d. Gaussian codebooks at the $M$ RUs and (mismatched) 
nearest-neighbor decoding at the $K$ UEs
\cite{scarlett2016dispersion, Choi2021FiniteSumrate}, the finite-blocklength
sum-rate can be written as 
\begin{equation}
    R_{{\rm p},\epsilon}(\mathbf{H},\mathbf{W})
    \hspace{-2pt}=\hspace{-2pt}
    \sum_{k=1}^{K}
    \bigg(
    \log_2(1+\gamma_{\rm p,k})
    -
    \sqrt{\frac{\mathcal{V}(\gamma_{\rm p,k})}{n}}
    \Phi^{-1}(1-\epsilon)
    \bigg),
    \label{eq:finite_sum_rate_precoding}
\end{equation}
where the signal-to-interference-plus-noise ratio (SINR) of UE $k$ is
\begin{equation}
    \gamma_{\rm p,k}
    =
    \frac{
    \left|
    \mathbf{h}_k^H\mathbf{w}_k
    \right|^2
    }{
    \sum_{\ell\neq k}
    \left|
    \mathbf{h}_k^H\mathbf{w}_{\ell}
    \right|^2
    +
    \sigma^2
    },
    \label{eq:precoding_sinr}
\end{equation}
$\mathcal{V}(\gamma_{\rm p,k})
=
\frac{2\gamma_{\rm p,k}}{1+\gamma_{\rm p,k}}\log_2^2(e)$ is the channel
dispersion factor, and $\Phi^{-1}(\cdot)$ is the inverse CDF of the standard
normal distribution.

\subsection{Design and Evaluation Details of AI-based Quantizer}
\label{quantizer_design_eval_appendix}

For fronthaul compression, we use a scalar uniform quantizer. Since
$\mathbf{x}_i\in\mathbb{C}^{NM}$, each RU quantizes $2N$ real scalar
components with $b_{\rm s}=\lfloor b_m/(2N)\rfloor$ bits per scalar. For an
input $u$, the quantization index is obtained as
\begin{equation}
    Q_{\rm uni}(u)
    =
    \mathrm{clip}
    \left(
        \left\lfloor
        \frac{
        \mathrm{clip}(u,u_{\min},u_{\max})-u_{\min}
        }{\Delta}
        \right\rceil,
        0,
        2^{b_{\rm s}}-1
    \right),
    \label{eq:scalar_uniform_quantizer}
\end{equation}
where $\lfloor\cdot\rceil$ denotes rounding to the nearest integer,
$\mathrm{clip}(x,a,b)=\min\{\max\{x,a\},b\}$, and
$\Delta=(u_{\max}-u_{\min})/(2^{b_{\rm s}}-1)$. The matched scalar dequantizer is
\begin{equation}
    Q_{\rm uni}^{-1}(a)
    =
    u_{\min}+\Delta a,
    \qquad
    a\in\{0,\ldots,2^{b_{\rm s}}-1\}.
    \label{eq:scalar_uniform_dequantizer}
\end{equation}

The CP-side AI-based quantization function in Sec.~III-B is
\begin{equation}
    Q(\mathbf{x}_i)
    =
    Q_{\rm uni}
    \big(
        g_Q(\mathbf{x}_i;\theta_Q)
    \big)
    \in \{0,1\}^{b}.
    \label{eq:ns_quantizer_function}
\end{equation}
The matched dequantization function at the $m$-th RU is
\begin{equation}
    Q_m^{-1}(\tilde{\mathbf{q}}_{i}^{m})
    =
    g_{Q,m}
    \left(
        Q_{m,\text{uni}}^{-1}(\tilde{\mathbf{q}}_{i}^{m});
        \theta_{Q,m}
    \right),
    \label{eq:ns_dequantizer_function}
\end{equation}
for $m=1,\ldots,M$. Although $g_{Q,m}$ is not the exact inverse of $g_Q$,
the two mappings are jointly trained to improve the end-to-end rate.

For rate evaluation, the dequantized signal is modeled by the Bussgang
decomposition \cite{demir2020bussgang} as
\begin{equation}
    \tilde{\mathbf{x}}_i
    \left(
        \theta_Q,\{\theta_{Q,m}\}_{m=1}^{M}
    \right)
    =
    \mathbf{A}_Q
    \left(
        \mathbf{W};
        \theta_Q,\{\theta_{Q,m}\}_{m=1}^{M}
    \right)
    \mathbf{x}_i
    +
    \mathbf{e}_i,
    \label{eq:ns_bussgang_decomp}
\end{equation}
where the Bussgang gain matrix is given by
\begin{equation}
    \mathbf{A}_Q
    \left(
        \mathbf{W};
        \theta_Q,\{\theta_{Q,m}\}_{m=1}^{M}
    \right)
    =
    \mathbb{E}
    \left[
    \tilde{\mathbf{x}}_i
    \mathbf{x}_i^H
    \right]
    \left(
    \mathbb{E}
    \left[
    \mathbf{x}_i\mathbf{x}_i^H
    \right]
    \right)^{-1},
    \label{eq:bussgang_gain_matrix}
\end{equation}
and the distortion covariance matrix is given by
\begin{equation}
    \boldsymbol{\Omega}_Q
    \left(
        \mathbf{W};
        \theta_Q,\{\theta_{Q,m}\}_{m=1}^{M}
    \right)
    =
    \mathbb{E}
    \left[
    \mathbf{e}_i\mathbf{e}_i^H
    \right].
    \label{eq:bussgang_distortion_covariance}
\end{equation}
Under $\mathbb{E}[\mathbf{s}_i\mathbf{s}_i^H]=\mathbf{I}_K$, the resulting
SINR of UE $k$ is
\begin{equation}
    \gamma_{{\rm q},k}
    =
    \frac{
    \left|
    \mathbf{h}_k^H
    \mathbf{A}_Q
    \mathbf{w}_k
    \right|^2
    }{
    \sum_{\ell\neq k}
    \left|
    \mathbf{h}_k^H
    \mathbf{A}_Q
    \mathbf{w}_{\ell}
    \right|^2
    +
    \mathbf{h}_k^H
    \boldsymbol{\Omega}_Q
    \mathbf{h}_k
    +
    \sigma^2
    }.
    \label{eq:quantized_sinr}
\end{equation}
The resulting Bussgang-type finite-blocklength rate surrogate can be written as 
\begin{equation}
\begin{aligned}
    &R_{{\rm q},\epsilon}
    \left(
        \mathbf{H},
        Q,
        \{Q_m^{-1}\}_{m=1}^{M}
    \right)
    =\\
   & \sum_{k=1}^{K}
    \left(
    \log_2(1+\gamma_{{\rm q},k})
    -
    \sqrt{\frac{\mathcal{V}(\gamma_{{\rm q},k})}{n}}
    \Phi^{-1}(1-\epsilon)
    \right),
    \end{aligned}
    \label{eq:finite_sum_rate_quantization}
\end{equation}
where $\mathcal{V}(\cdot)$ and $\Phi^{-1}(\cdot)$ are defined in
\eqref{eq:finite_sum_rate_precoding}. In (\ref{eq:finite_sum_rate_quantization}), we have considered the worst case (in terms of the first-order asymptotics) uncorrelated additive noise theorem \cite{hassibi2003much} to obtain a tractable rate surrogate. Furthermore, throughout our whole work, we always approximate the matrices  (\ref{eq:bussgang_gain_matrix})--(\ref{eq:bussgang_distortion_covariance}) with $N_\text{MC}=100$ Monte Carlo samples whose randomness comes from the random input symbol vector    $\mathbf{s} \sim P_\mathbf{s}$.  

Specifically, denoting as $\{ \mathbf{s}^{(r)} \}_{r=1}^{N_\text{MC}}$ the $N_\text{MC}$ random input symbol vectors with  $\mathbf{s}^{(r)} \overset{\text{i.i.d.}}{\sim} P_\textbf{s}$ for $r=1,...,N_\text{MC}$, we can accordingly  express the corresponding (unquantized) precoded symbol vectors as $\mathbf{x}^{(r)}=\mathbf{W}\mathbf{s}^{(r)}$. Further denoting as $\tilde{\mathbf{x}}^{(r)}$ the  corresponding dequantized symbol vector, the Bussgang gain and distortion covariance matrices can be estimated via Monte Carlo sampling
\cite{Ren2022MonteCarlo} as
\begin{equation}
    \widehat{\mathbf{A}}_Q
    \hspace{-2pt}=\hspace{-2pt}
    \bigg(
    \frac{1}{N_{\rm MC}}
    \sum_{r=1}^{N_{\rm MC}}
    \tilde{\mathbf{x}}^{(r)}(\mathbf{x}^{(r)})^H
    \bigg)
    \bigg(
    \frac{1}{N_{\rm MC}}
    \sum_{r=1}^{N_{\rm MC}}
    \mathbf{x}^{(r)}(\mathbf{x}^{(r)})^H
    \bigg)^{-1},
    \label{eq:mc_gain}
\end{equation}
and
\begin{equation}
    \widehat{\boldsymbol{\Omega}}_Q
    =
    \frac{1}{N_{\rm MC}}
    \sum_{r=1}^{N_{\rm MC}}
    \mathbf{e}^{(r)}(\mathbf{e}^{(r)})^H,
        \label{eq:mc_distortion_covariance}
\end{equation}
where $\mathbf{e}^{(r)}
    =
    \tilde{\mathbf{x}}^{(r)}
    -
    \widehat{\mathbf{A}}_Q\mathbf{x}^{(r)} $.




\subsection{Implementation and Training Details of the Test-Time Scalable Precoder}
\label{appendix:tts_precoder_details}

We train $\theta_W'$ with the aim at  $G_W(\cdot;\theta_W')$ generating a candidate
set containing high-rate precoders  after WMMSE refinement. Under the fixed
ordered sampling rule, we consider
\begin{equation}
\begin{aligned}
    \max_{\theta_W'}\quad
    &\mathbb{E}_{\mathbf{H}}
    \left[
    \max_{j\in\{1,\ldots,J\}}
    R_{{\rm p},\epsilon}(\mathbf{H},\mathbf{W}_j)
    \right]
    \nonumber\\
    \text{s.t.}\quad
    &\mathbf{W}_j
    =
    \mathrm{WMMSE}_{I_W}
    \big(
        \mathbf{W}_{0,j},\mathbf{H}
    \big),
    \nonumber\\
    &\mathbf{W}_{0,j}\overset{\rm i.i.d.}{\sim}G_W(\mathbf{H};\theta_W'),
    \qquad j=1,\ldots,J .
    \label{eq:tts_ai_wmmse_training_problem}
    \end{aligned}
\end{equation}
This objective promotes a high $R_{{\rm p},\epsilon}$ for the best refined
precoder.

In practice, $G_W(\cdot;\theta_W')$ augments the deterministic initializer
$g_W(\cdot;\theta_W)$ in Sec.~\ref{subsec:non_scalable_ai_precoder} with an
auxiliary random input:
\begin{equation}
    \mathbf{W}_{0,j}
    =
    g_W(\mathbf{H},\boldsymbol{\xi}_j;\theta_W'),
    \qquad
    \boldsymbol{\xi}_j\overset{\rm i.i.d.}{\sim}P_{\boldsymbol{\xi}},
\end{equation}
with the output normalized or projected onto $\mathcal{W}$ if needed. 


During empirical training, the transmission duration is randomly sampled within $T$, and the corresponding
blocklength is used for rate evaluation. The objective in
\eqref{eq:tts_ai_wmmse_training_problem} is optimized by an empirical loss. Given a mini-batch $\mathcal{B}$, we generate the first
$J^{\rm tr}\leq J$ candidates from the fixed ordered seed sequence and replace
the maximum by the log-sum-exp approximation:
\begin{equation}
    \widehat{\mathcal{L}}_W(\theta_W')
    =
    -
    \frac{1}{|\mathcal{B}|}
    \sum_{\mathbf{H}\in\mathcal{B}}
    \eta_W
    \log
    \sum_{j=1}^{J^{\rm tr}}
    \exp
    \left(
    \frac{
    R_{{\rm p},\epsilon}
    }{\eta_W}
    \right),
    \label{eq:tts_precoder_training_loss}
\end{equation}
where $\eta_W>0$ is a temperature parameter, and the $j$-th candidate precoder is given by \eqref{eq:tts_precoder_candidate}. As $\eta_W$ decreases, the loss approaches the negative maximum over the candidates.

\subsection{Implementation and Training Details of the Test-Time Scalable Quantizer}
\label{appendix:tts_quantizer_details}

We train $\theta_Q'$ and $\{\theta_{Q,m}'\}_{m=1}^{M}$ with the aim at $G_Q(\theta_Q')$ and $\{G_{Q,m}(\theta'_{Q,m})\}_{m=1}^M$ generating a candidate
set  containing high-rate quantization-dequantization designs. Under the fixed
ordered sampling rule, we consider
\begin{equation}
\begin{array}{@{}l@{\quad}l@{}}
\displaystyle
\max_{\theta_Q',\{\theta_{Q,m}'\}_{m=1}^{M}}
&
\displaystyle \hspace{-10pt}
\mathbb{E}_{\mathbf{H}}
\left[
\max_{j\in\{1,\ldots,J\}}
R_{{\rm q},\epsilon}
\left(
    \mathbf{H},
    Q_j,
    \{Q_{j,m}^{-1}\}_{m=1}^{M}
\right)
\right]
\\[1ex]
\multicolumn{2}{@{}l@{}}{
\displaystyle
\text{s.t.}\quad
Q_j(\cdot)
=
Q_{\rm uni}\circ g_Q(\cdot;\theta_{Q,j}),
}
\\[1ex]
\multicolumn{2}{@{}l@{}}{
\displaystyle
Q_{j,m}^{-1}(\cdot)
=
g_{Q,m}(\cdot;\theta_{Q,m,j})\circ
Q_{m,{\rm uni}}^{-1}(\cdot),
}
\\[1ex]
\multicolumn{2}{@{}l@{}}{
\displaystyle
\theta_{Q,j}
\overset{\rm i.i.d.}{\sim}
G_Q(\theta_Q'),
\theta_{Q,m,j}
\overset{\rm i.i.d.}{\sim}
G_{Q,m}(\theta_{Q,m}'),}
\end{array}
\label{eq:tts_ai_quantizer_training_problem}
\end{equation}
for $m=1,\ldots,M$ and $j=1,\ldots,J$. This objective promotes a high $R_{{\rm q},\epsilon}$ for the best matched
tuple.

The parameters $\theta_Q'$ and $\theta_{Q,m}'$ include the base parameters 
$\theta_Q$ and $\theta_{Q,m}$, respectively, and the stochastic-perturbation
parameters. We specifically consider the following implementation via random noise injection:
\begin{equation}
    \theta_{Q,j}
    =
    \theta_Q+\boldsymbol{\varepsilon}_j,
    \qquad
    \theta_{Q,m,j}
    =
    \theta_{Q,m}+\boldsymbol{\varepsilon}_{m,j},
\end{equation}
where
$(\boldsymbol{\varepsilon}_j,\{\boldsymbol{\varepsilon}_{m,j}\}_{m=1}^{M})$
is generated from a fixed ordered random seed sequence. The perturbation is applied to the quantization-dequantization functions,
not each input symbol, keeping the CP-side quantizer and RU-side dequantizers
matched for all symbols of the same candidate.

Similarly, $R_{{\rm q},\epsilon}$ is evaluated using transmission durations
randomly sampled within $T$. The objective in
\eqref{eq:tts_ai_quantizer_training_problem} is optimized by an empirical loss. Given a mini-batch
$\mathcal{B}=\{(\mathbf{H}_b,\mathbf{W}_b)\}_{b=1}^{|\mathcal{B}|}$, we
generate the first $J^{\rm tr}\leq J$ tuples from the fixed ordered seed
sequence and use the log-sum-exp approximation:
\begin{equation}
\begin{aligned}
&\widehat{\mathcal{L}}_Q
\left(
    \theta_Q',
    \{\theta_{Q,m}'\}_{m=1}^{M}
\right)
= \\
&
-
\frac{1}{|\mathcal{B}|}
\sum_{(\mathbf{H}_b,\mathbf{W}_b)\in\mathcal{B}}
\eta_Q
\log
\sum_{j=1}^{J^{\rm tr}}
\exp
\left(
\frac{
R_{{\rm q},\epsilon}}{\eta_Q}
\right),
\end{aligned}
\label{eq:tts_quantizer_training_loss}
\end{equation}
where $\eta_Q>0$ is a temperature parameter and the $j$-th candidate tuple is
obtained from \eqref{eq:tts_quantizer_candidate_tuple}. As $\eta_Q$ decreases,
the loss approaches the negative maximum over the candidate
quantization-dequantization tuples.

    \subsection{Test-time Scalability of the Precoding and Quantization Module under Mismatched CSI Conditions}
\label{appendix_outdated}
    

        \begin{figure}[t]
        \centering
        \includegraphics[width=.6\columnwidth]{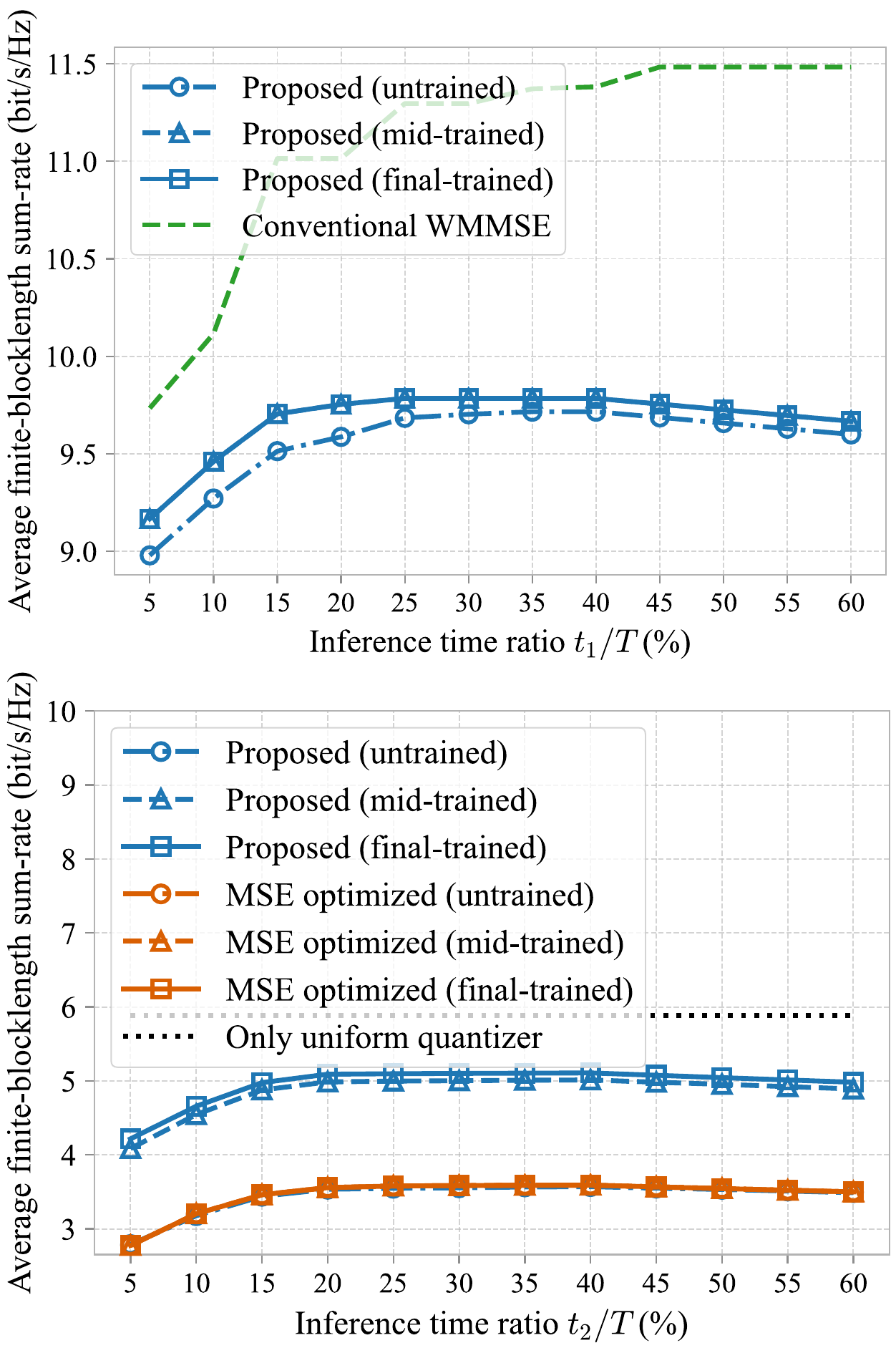}
        \caption{Average finite-blocklength sum-rate under outdated CSI conditions with $M=4$ RUs equipped with $N=64$ antennas, and $K=4$ single-antenna UEs: (a) as a function of the precoding function inference time ratio $t_1/T$; (b) as a function of the quantization function inference time ratio $t_2/T$.}
        \label{result_outdated_inference}
    \end{figure}
    To examine the generalization capability of the stochastic modules, we consider a mismatched CSI scenario: Unlike
Figs.~\ref{result_precoder_inference} and \ref{result_quantizer_inference},
where training and testing use the same outdoor environment, the models are
trained in the original scenario and tested in a different deployment
environment with substantially different channel characteristics. As shown in
Fig.~\ref{result_outdated_inference}, the proposed data-driven methods suffer
noticeable degradation under this mismatch, indicating their dependence on
the relevance of training data to the deployment environment. 


    \bibliographystyle{IEEEtran}
    \bibliography{ref}

    \end{document}